\documentclass{article}
\usepackage{graphicx} % Required for inserting images
\usepackage[title]{appendix}
\usepackage[numbers]{natbib}
\usepackage{amsmath}
\usepackage{geometry}
\usepackage{booktabs}
\usepackage{caption}
\usepackage{subcaption}
\usepackage{hyperref}
\usepackage{listings}
\hypersetup{
    colorlinks=true,
    linkcolor=blue,
    filecolor=magenta,      
    urlcolor=cyan,
    pdftitle={Overleaf Example},
    pdfpagemode=FullScreen,
    }
\usepackage{titlesec}
\usepackage{orcidlink}
\usepackage{authblk}

\newcommand{\beginsupplement}{%
    \setcounter{table}{0}
    \renewcommand{\thetable}{S\arabic{table}}%
    \setcounter{figure}{0}
    \renewcommand{\thefigure}{S\arabic{figure}}%
    \setcounter{section}{0}
    \renewcommand{\thesection}{S\arabic{section}}%
}

\title{When Active Learning Fails, Uncalibrated Out of Distribution Uncertainty Quantification Might Be the Problem}
\author[1]{Ashley S. Dale\orcidlink{0000-0001-8233-5258}\footnote{ashley.dale@utoronto.ca}} 
\author[2]{Kangming Li\orcidlink{0000-0003-4471-8527}}
\author[3]{Brian DeCost\orcidlink{0000-0002-3459-5888}}
\author[1]{Hao Wan\orcidlink{0000-0002-7489-3433}} 
\author[1]{\\ Yuchen Han\orcidlink{0009-0003-6579-3929}}
\author[4]{Yao Fehlis\orcidlink{0009-0004-1300-7364}}
\author[1, 5, 6, 7]{Jason Hattrick-Simpers\orcidlink{0000-0003-2937-3188}\footnote{jason.hattrick.simpers@utoronto.ca}}

\affil[1]{\emph{\small Department of Materials Science and Engineering, University of Toronto, 27 King’s College Cir, Toronto, ON, Canada}}

\affil[2]{\emph{\small Physical Science and Engineering Division (PSE), King Abdullah University of Science and Technology (KAUST), Thuwal 23955-6900, Saudi Arabia}}

\affil[3]{\emph{\small Material Measurement Laboratory, National Institute of Standards and Technology, 100 Bureau Dr, Gaithersburg, MD, USA.}}

\affil[4]{\emph{\small Artificial, Inc.}}
 
\affil[5]{\emph{\small Acceleration Consortium, University of Toronto, 27 King’s College Cir, Toronto, ON, Canada.}}

\affil[6]{\emph{\small Vector Institute for Artificial Intelligence, 661 University Ave, Toronto, ON, Canada.}}

\affil[7]{\emph{\small Schwartz Reisman Institute for Technology and Society, 101 College St, Toronto, ON, Canada.}}

\date{}
\begin{document}

\maketitle

\begin{abstract}

Efficiently and meaningfully estimating prediction uncertainty is important for exploration in active learning campaigns in materials discovery, where samples with high uncertainty are interpreted as containing information missing from the model. In this work, the effect of different uncertainty estimation and calibration methods are evaluated for active learning when using ensembles of ALIGNN, eXtreme Gradient Boost, Random Forest, and Neural Network model architectures. We compare uncertainty estimates from ALIGNN deep ensembles to loss landscape uncertainty estimates obtained for solubility, bandgap, and formation energy prediction tasks. We then evaluate how the quality of the uncertainty estimate impacts an active learning campaign that seeks model generalization to out-of-distribution data. Uncertainty calibration methods were found to variably generalize from in-domain data to out-of-domain data. Furthermore, calibrated uncertainties were generally unsuccessful in reducing the amount of data required by a model to improve during an active learning campaign on out-of-distribution data when compared to random sampling and uncalibrated uncertainties. The impact of poor-quality uncertainty persists for random forest and eXtreme Gradient Boosting models trained on the same data for the same tasks, indicating that this is at least partially intrinsic to the data and not due to model capacity alone. Analysis of the target, in-distribution uncertainty, out-of-distribution uncertainty, and training residual distributions suggest that future work focus on understanding empirical uncertainties in the feature input space for cases where ensemble prediction variances do not accurately capture the missing information required for the model to generalize.

\end{abstract}

 % \tableofcontents
\section*{Introduction}
Modern materials discovery~\cite{xu2023small, pilania2021machine, chen2024accelerating, chen2024machine,shahzad2024accelerating, mulukutla2024illustrating} and drug discovery~\cite{gupta2021artificial, edfeldt2024data} commonly leverage uncertainty estimates in methods such as active learning~\cite{wang2022benchmarking}. In an active learning campaign, a model is trained using a small initial dataset, then data is iteratively added using a pre-defined acquisition policy. A common choice of acquisition policy chooses samples with high-uncertainty predictions to add to the training dataset; this approach to active learning is known as ``exploration''~\cite{wang2022benchmarking}. In sparse-data regimes, e.g. an experiment where only a small number of samples can be created and characterized, a robust uncertainty estimate may differentiate between a model that is predictive and generalizable and one that is not~\cite{choudhary2022recent}. 

In many cases, the uncertainty distribution used for active learning is presumed to be reliable based on previous knowledge of the model/data or mathematical modeling, and explicit calibration of the uncertainty is omitted. For example, random forest (RF) models have been evaluated and shown to have intrinsically calibrated uncertainty estimates~\cite{shaker2025random}. Bayesian neural networks are designed a priori to have calibrated uncertainty estimates~\cite{mitros2019validity}. A successful active discovery campaign that leverages exploration is empirical evidence that the uncertainty estimates are meaningful~\cite{li2023exploiting, kusne2020fly}, especially when exploration is more successful than random selection. However, in the absence of explicit calibration validating the uncertainty estimates, there is a missed opportunity to understand and quantify subtle differences between the true data distribution and the uncertainty distribution learned by the model.

Although some uncertainty estimates are useful in an uncalibrated state, other models suffer from miscalibration~\cite{zhang2021leveraging, varivoda2023materials}, where the predicted uncertainty distribution is not consistent with the model accuracy and error distribution on testing set data. There is disagreement as to whether a particular method produces a reliable uncertainty estimate or not. For example,~\cite{shaker2025random} report that calibrating the uncertainty of random forests may be unnecessary or detrimental, while~\cite{dankowski2016calibrating} report that calibration is necessary. Deep ensemble approaches requires careful calibration to achieve a reliable uncertainty estimate, so that the quality of the uncertainty estimate may depend on the order that calibration method steps are applied~\cite{rahaman2021uncertainty, wenzel2020hyperparameter}. A held-out test dataset is used in these cases to validate the uncertainty estimates after calibration, just as a held-out test dataset is used to validate model accuracy on unseen data. When the unseen data is in-distribution (ID) with respect to the training data, this is a reasonable approach.  However, when the unseen data is out-of-distribution (OOD) and the researcher is pursuing generalization, we find that this approach may be invalid. Addressing potential miscalibration between ID and OOD data is therefore important to active learning methodologies, where the effect of acquiring non-optimal data may prevent the model from generalizing to unseen data or incur additional training costs. 

A good understanding of OOD uncertainty estimation is also relevant to a broader understanding of model generalizability and interpretability beyond active learning tasks. High predictive uncertainty on the training data suggests the model is underconfident and the predicted uncertainty is larger than the actual errors, while uncertainty estimates for test data that are smaller than the training data indicate overconfidence~\cite{tran2020methods}. When underconfident or overconfident uncertainty estimates are reported as confidence intervals or error bars for predictions~\cite{tran2020methods}, this can confuse interpretation of the final result. Reliably quantifying the uncertainty, even if it will not be further leveraged in an active learning campaign, is therefore a useful method of evaluating model performance beyond accuracy.

The main contribution of this work is a study of uncertainty sources, calibration methods and the impact on the quality of information represented by the calibrated uncertainty estimate for three different tasks. To generate uncertainty estimates from a single pre-trained model, we sample the loss landscape to generate an ensemble of models and demonstrate that such methods behave comparably to uncertainty estimates from ensembles created by retraining the model using different random seeds. The uncertainty estimates are then calibrated using two different methods, including a novel approach using a neural network. We find that an initial, uncalibrated uncertainty estimate better captures the true uncertainty of OOD data, and that the use of uncertainty calibration models optimized for ID uncertainty distributions can degrade OOD uncertainty estimates. If the model has overfit to the training distribution, using ID data to perform uncertainty calibration decreases the generalizability of the model's uncertainty estimates. We extend the analysis from neural networks to random forest and XGBoost models trained for the same task, and observe similar trends for OOD uncertainty. As a result, we consider the increased miscalibration of OOD data a property of the dataset and how the model is trained, rather than a property of any one model architecture. Finally, we show that the impact of uncertainty calibration is subtle, and cannot be understood by analyzing the prediction space alone. To summarize, we observe that \textit{for OOD data, inaccurate predictions correlate with uncalibrated uncertainties and common uncertainty calibration methods may worsen the situation.} 

\begin{figure}[!t]
  \centering
  \includegraphics[width=\linewidth]{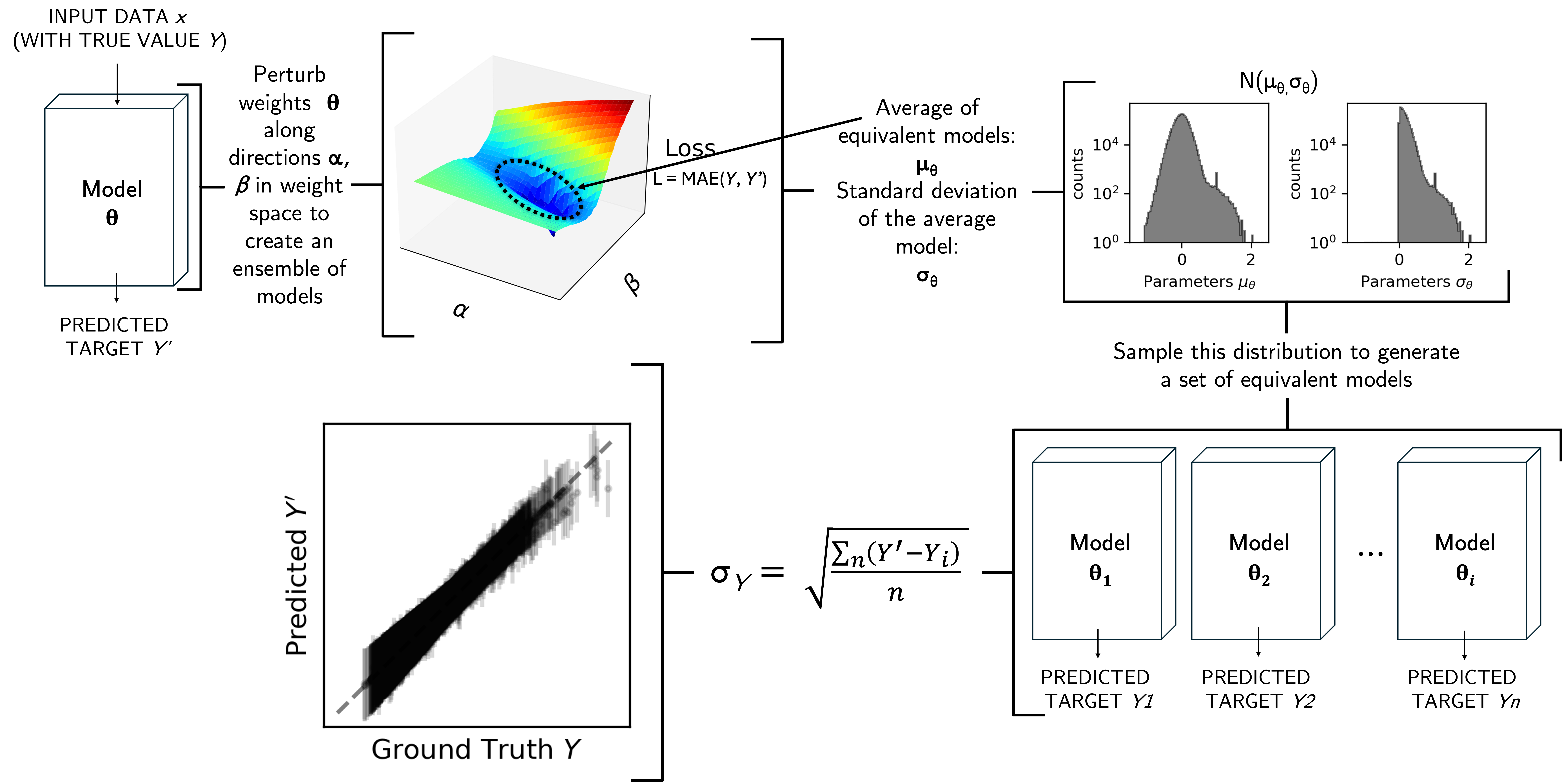}
  \caption{Example use of a loss landscape to generate uncertainty predictions. After training, a model's weights $\theta$ are perturbed along orthogonal directions $\alpha, \beta$ in the weight space. The perturbed model's performance is quantified using the model loss, forming the loss landscape. Models with similar performance in the loss landscape are reserved. The average $\mu_\theta$ and  standard deviation $\sigma_\theta$ of each parameter are reserved to create an independent Gaussian approximation of the posterior distribution over model parameters $\mathcal{N} \left( \mu_\theta, \sigma_\theta \right)$. Additional models are sampled from this distribution, and used to generate a distribution of predictions with uncertainty $\sigma_{Y}$.}
  \label{fig:summary}
\end{figure}

% \clearpage
% \input{sec/2_background}
% % \clearpage
\section*{Results}
\label{sec:results}

\subsection*{Evaluation of uncertainty estimates}

We evaluate the reliability of uncertainty estimates and calibration methods in an active learning context. To accomplish this, we partition data as in-distribution (ID) and out-of-distribution (OOD) following~\cite{li2025probing} or a clustering approach (see ``Methods''). The ID data is then partitioned again into ID-train and ID-test data for use in model training. The OOD data is reserved in its entirety for calibration assessment. The active learning experiments further divide the OOD data into OOD-train for the query pool and OOD-test for active learning performance evaluation.

Uncertainty estimates are obtained using two methods: First, from an ensemble of models trained with the same ID-train dataset but with different random seeds following common practice~\cite{rahaman2021uncertainty}; uncertainty estimates from the random-ensemble are the standard deviation of predictions from five models. Second, from a loss landscape (LL) approach~\cite{ravishankar2022stochastic, maddox2019simple} that uses a single model and the training dataset for that model, then interpolates the weights along the two Hessian eigenvectors associated with the min/max Hessian eigenvalues to form a plane. As shown in Figure \ref{fig:summary}, the LL-ensemble of models is created by sampling around the minima of the LL, using the sampled models to create a normal distribution of candidate models $\mathcal{N}\left(\mu_\theta, \sigma_\theta\right)$, then sampling this distribution repeatedly to obtain a diverse set of models and predictions. Uncertainty estimates from the LL-ensemble are the standard deviation of predictions from 500 models generated from the distribution $\mathcal{N}\left(\mu_\theta,\sigma_\theta\right)$ (see ``Methods''). For a well-calibrated uncertainty estimate, large prediction error is correlated with high uncertainty, so that the probability distribution of the errors on the test set is consistent with the predicted uncertainty distribution. For example, 10~\% of the test data is expected to lie outside of the 90~\% confidence interval around the mean prediction. Enforcing this relationship motivates the uncertainty calibration adjustment step for poorly-calibrated models.

After uncertainty estimates are created for each sample, we implement two methods of uncertainty calibration: linear adjustment using calibration factors~\cite{palmer2022calibration}, and the direct modeling of the posterior using a neural network, similar to the method proposed in~\cite{dey2025instancewise}. In each case, the uncertainty distribution is calibrated against the ID-test residual distribution. This ensures that each ID-test uncertainty interval contains the true value for that sample with the appropriate probability. The calibration model tuned to the ID-test residual distribution is then used to calibrate the OOD uncertainty distribution. 

We then evaluate the quality of uncertainty estimate in three active learning tasks that predict the formation energy, bandgap, and solubility of different materials. The active learning method implements an acquisition policy that selects OOD-train samples with the largest uncertainty and adds them to the ID-train dataset (see ``Methods"). If the uncertainty estimate does not represent the true uncertainty of the model, then adding that sample to the dataset may not improve the model's performance over random selection. 

Different model architectures were used to ensure generalizable analysis. For the formation energy and bandgap prediction tasks, we implement Atomistic Line Graph Neural Network (ALIGNN)~\cite{choudhary2021atomistic} architectures, random forest (RF)~\cite{breiman2001random} models, and XGBoost (XGB)~\cite{chen2016xgboost} models. ALIGNN models are shown to poorly perform on some of the out-of-distribution tasks, and this poor predictive ability is dependent on the property chosen for the prediction task and the chemistry available to the model during training \cite{li2023exploiting, li2025probing}. These models were trained using the JARVIS 3D DFT dataset~\cite{choudhary2020joint} where the ID data is the entire dataset omitting chemistries containing either iron (Fe) for case one, or fluorine (F) for case two. The OOD dataset consists of samples which contain the omitted element. Fluorine was selected due to its previous identification as an element to which an ALIGNN model would not generalize to unseen~\cite{li2025probing}, and therefore expected to have high uncertainty as an OOD dataset. Iron was selected as a control due to the ALIGNN model's ability to generalize to it unseen~\cite{li2025probing}. For the solubility prediction task, the Estimated Solubility (ESOL) dataset \cite{delaney2004esol} from the MoleculeNet benchmark \cite{wu2018moleculenet} is a widely used for evaluating machine learning models in computational chemistry, specifically for predicting aqueous solubility of small molecules. A simple multilayer perceptron model was trained using ID training data identified using a clustering technique. Similar trends regarding ID and OOD data uncertainties for the different model architectures strengthen the conclusion that the relationship between ID and OOD uncertainty estimates for active learning is not strictly a property of the model architecture or the information to which the model has access, but rather a combination of the two. 

\subsection*{Loss landscape analysis}
\begin{figure}[!h]
    \centering
    \includegraphics[width=\linewidth]{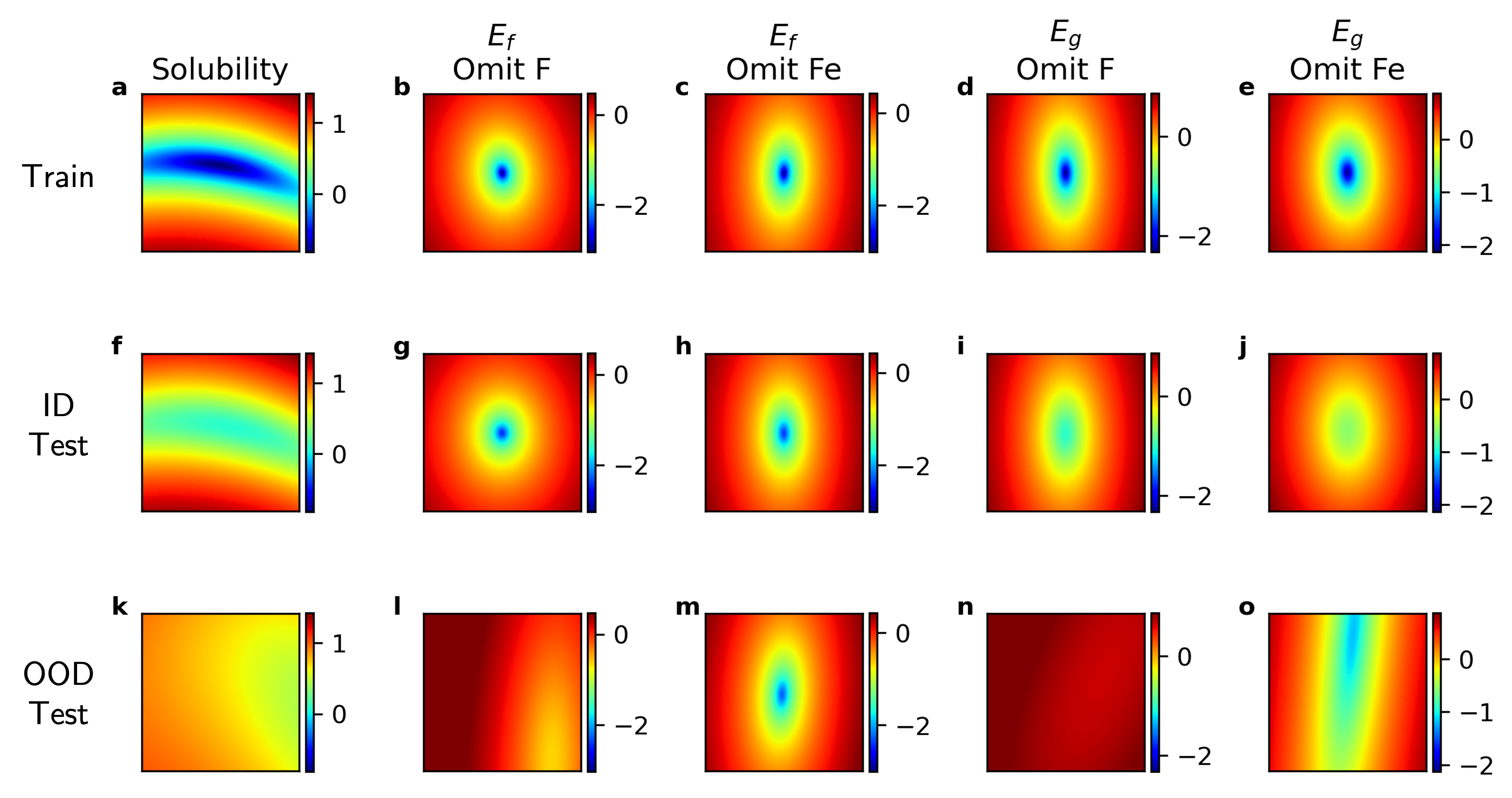}
    \caption{Loss Landscapes. Each column represents a different data study: Column 1 (a, f, k) is the solubility prediction task, column 2 (b, g, l) the formation energy $E_f$ task with fluorine (F) omitted from the training data, column 3 (c, h, m) is the formation energy task $E_f$ with iron (Fe) omitted from the training data, column 4 (d, i, n) is the bandgap $E_g$ prediction task with F omitted from the training data, and column 5 (e, j, o) is the bandgap $E_g$ prediction task with Fe omitted from the training data. Row 1: loss landscapes from training data. Row 2: loss landscapes from ID test data. Row 3: loss landscapes from OOD data. The colorbar visualizes the log loss for each task, while the x-axis and y-axis are two directions in the weight space of the original model.}
    \label{fig:loss_landscapes}
\end{figure}

Given a well-performing model on the training data, the loss landscape (LL) analysis may proceed. Each LL plot is the prediction error for a single model trained using  ID data and perturbed with Hessian eigenvectors for the trained model. The LLs associated with the solubility task, formation energy $E_f$ tasks and bandgap $E_g$ prediction tasks are shown in Figure~\ref{fig:loss_landscapes}. In this figure, the LLs generated using ID data samples have distinct geometric features from LLs generated using OOD data, e.g.~the minimum of the ID LL is not coincident with or has different magnitudes than the minimum of the OOD LL. Each pixel in the LL represents the prediction generated by a model with perturbed weights; the data are held constant for each interpolation step. The original model weights are located in the center of each LL. Table~\ref{tab:pred_val} reports the mean absolute error (MAE) for these original models losses for the Training, ID Test, and OOD Test sets.

The first row of LLs in Figure~\ref{fig:loss_landscapes} visualize the training data log loss, the second row of LLs visualizes the ID-test data log loss, and the third row of LLs visualize the OOD data log-loss. The dark blue center represents the minima of each landscape, while the ellipticity of each landscape is due to the selection of min/max eigenvalued eigenvectors from the Hessian of the model. In Figure~\ref{fig:loss_landscapes}(a), the LL has the form of a pseudo-saddle point, indicating a lack of stability during model convergence compared to (b, c, d, e). In Figure~\ref{fig:loss_landscapes}(f, i, j), the performance gap between the training data and the ID test data is visualized as the disappearance of the dark blue minima. The various predictive performances on the OOD data vary widely, from minimal performance gap  in (m), to moderate performance degradation (k, o), and finally a complete failure to generalize (l, n).

The general trend observed in the LLs is that the ID-test data in Figure~\ref{fig:loss_landscapes}(f-j) tends to have slightly worse performance (higher loss values) compared to the training data. However, performance on the OOD-test data as visualized in Figure~\ref{fig:loss_landscapes}(k, l, n, o) is markedly different compared to the training data. Importantly, the OOD LL in (k, l, n, o) indicate different types of degradation behaviors. In Figure~\ref{fig:loss_landscapes}(k) and (o), the minima for the OOD data is displaced away from the center, but visible, suggesting that a different training approach may allow the model to generalize~\cite{garipov2018loss}. In Figure~\ref{fig:loss_landscapes}, the model loss has increased (performance has decreased) so that a minima is no longer observable---both of these cases occur when fluorine has been omitted from the training data. 

The differences in the LL between the first row and the third row indicate that the uncertainties for the ID-test data and the OOD data should have distinct distributions. The loss for the OOD data is not constant at the location of the minima for the ID data. The average loss correlates with the residual error for each dataset, and by definition a calibrated uncertainty distribution will correlate with the distribution of residuals. Since the OOD loss around the ID minima does not co-vary with the ID loss, this suggests that the ID and OOD data residual distributions will also not co-vary, and that the uncertainty distributions will be distinct. This is further explored in the following section.

\begin{table}[h]
    \centering
    \caption{Performance metric $R^2$ for single ALIGNN models trained for predicting two different targets (bandgap and formation energy) using two different omitted elements (F and Fe), and for a multi-layer perceptron (MLP) model trained for predicting solubility.}
    \begin{tabular}{lp{0.2\linewidth}p{0.2\linewidth}p{0.2\linewidth}}
         \textbf{Model} & \textbf{Train MAE ($R^2$)} & \textbf{ID Test MAE ($R^2$)} & \textbf{OOD MAE ($R^2$)}\\
        \hline
        Solubility &1.768 (-0.076) &0.059 (0.999) &1.671 (-2.956)\\
         $E_f$-F & 0.015 (0.999) & 0.036 (0.996) & 1.091 (-0.641)\\
        $E_f$-Fe & 0.016 (0.999) & 0.037 (0.993)& 0.064 (0.992)\\
         $E_g$-F & 0.041 (0.995) & 0.137 (0.911) & 1.244 (-7.065)\\
         $E_g$-Fe & 0.050 (0.994) & 0.155 (0.915) & 0.171 (0.033)\\
    \end{tabular}
    \label{tab:pred_val}
\end{table}

\subsection*{Uncertainty quantification and calibration}
\begin{figure}[!h]
\begin{subfigure}[b]{\textwidth}
    \includegraphics[width=\linewidth]{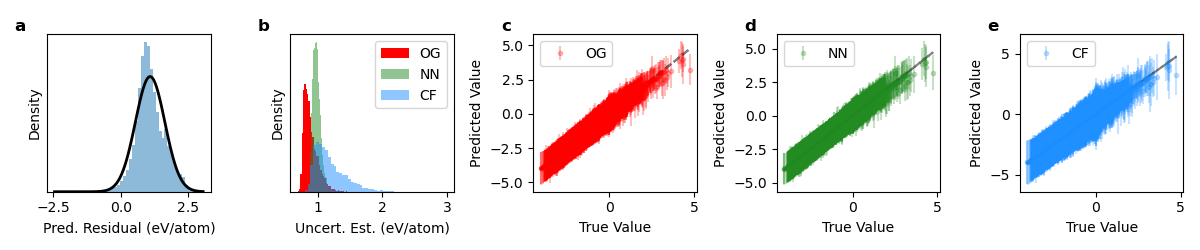}
    \label{fig:eform_F_id_recal_ll}
\end{subfigure}
\begin{subfigure}[b]{\textwidth}
    \includegraphics[width=\linewidth]{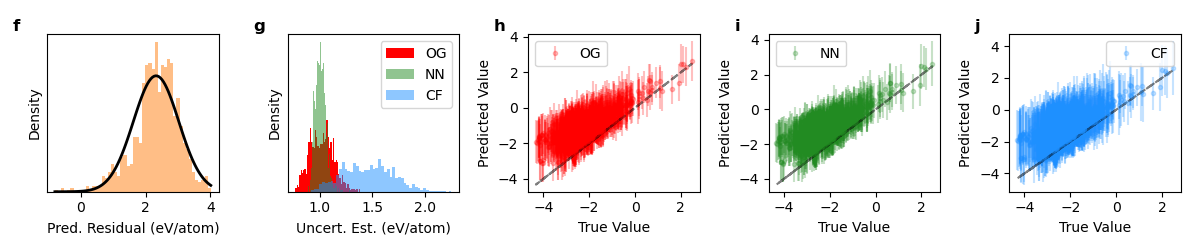}
    \label{fig:eform_F_ood_recal_ll}
\end{subfigure}
\begin{subfigure}[b]{\textwidth}
    \includegraphics[width=\linewidth]{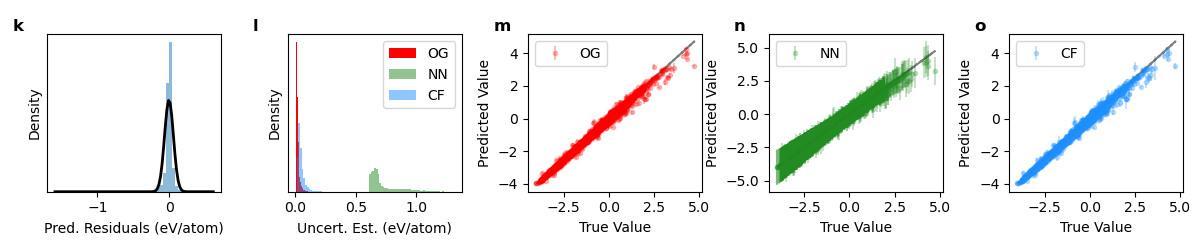}
    \label{fig:eform_F_id_recal_ensem}
\end{subfigure}
\begin{subfigure}[b]{\textwidth}
    \includegraphics[width=\linewidth]{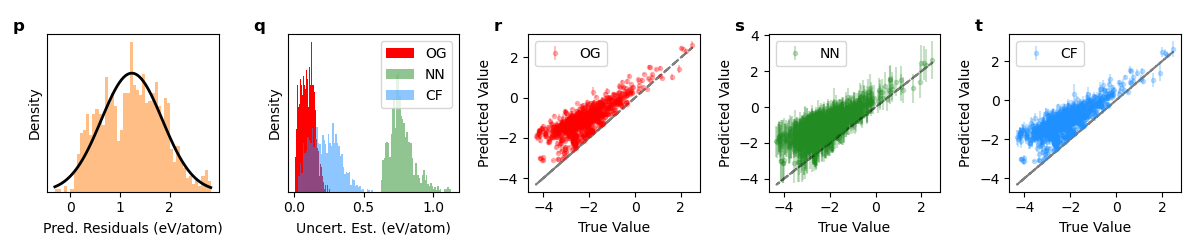}
    \label{fig:eform_F_ood_recal_ensem}
\end{subfigure}
\caption{Comparison of uncertainty and predictions from LL-ensemble (a-j) and random-ensemble (k-t) for the formation energy $E_f$ task when F-containing compounds are omitted from the training data. Column 1 is the prediction residuals on the in-distribution (blue) and out-of-distribution (orange) data. Column 2 is the uncertainty estimates for loss landscape (b, g) and ensemble (i, q) methods; the original (OG) uncertainty distribution is shown in red, the neural network (NN) calibration method in green, and the calibration factor (CF) method shown in blue. Columns 3-5 show the uncertainty distributions of column 2 as error bars on parity plots.}
\label{fig:eform_F_uncert_hist_parity}
\end{figure}

Next, we compare uncertainty estimates from an LL-ensemble to a random-ensemble created by retraining the model with different random seeds, and investigate the effect of uncertainty calibration on ID and OOD uncertainty estimates.  The uncertainties of the OOD data are calibrated against the distribution of residual values for the ID-test data, visualized in Figure~\ref{fig:eform_F_uncert_hist_parity} column one for the formation energy $E_f$ prediction task when the model is trained omitting fluorine (F) for LL-ensemble uncertainty estimates (a-j) and random-ensemble uncertainty estimates (k-t). Column 2 is the uncertainty estimates for loss landscape (b, g) and ensemble (i, q) methods; the original (OG) uncertainty distribution is shown in red, the neural network (NN) calibration method in green, and the calibration factor (CF) method shown in blue. Columns 3-5 show the uncertainty distributions of column 2 as error bars on parity plots. Equivalent figures for the remaining property studies may be found in the Supplement.

In column one (a, f, k, p) of Figure~\ref{fig:eform_F_uncert_hist_parity}, the residuals of the mean ensemble prediction for the ID-test data are blue (a, k), while the OOD data residuals are orange (f, p). Rows one and two are results from the LL-ensemble approach, and rows three and four are from the random-ensemble approach. The right-shifted residual distributions in rows one and two compared to rows three and four are due to a permissive sampling strategy in the LL; models with average training loss $\mathcal{L} <0.1$ eV/atom were included in the $\mathcal{N}\left(\mu_\theta,\sigma_\theta\right)$ distribution of models for sampling the LL-ensemble. The LL sampling threshold of $\mathcal{L} <0.1$ eV/atom is $\approx1.5$ orders of magnitude larger than the loss of the original model.  In contrast, the random-ensemble models were each trained to an equivalent loss value, resulting in smaller residuals than the LL-ensemble estimate.  

The ID-test residuals in Figure~\ref{fig:eform_F_uncert_hist_parity} resemble a Gaussian or Lorentzian distribution, and this is confirmed by the Gaussian fit applied to the distribution (visualized as a black line). The OOD residuals in Figure~\ref{fig:eform_F_uncert_hist_parity}(f, p) suggest bimodal distributions. Considering the mismatch between training/ID-test LL and the OOD LL in Figure~\ref{fig:loss_landscapes}, this is a second indication that uncertainty distributions for the ID data are distinct from the uncertainty distributions for the OOD data, since we desire that the uncertainty distribution follow the residual distribution.

The ID-test original (OG, red) uncertainty estimates are calibrated using the neural network (NN, green) and calibration factors (CF, blue) methods. These values are shown in column two of Figure~\ref{fig:eform_F_uncert_hist_parity} as histograms, and as error bars on the parity plots in columns three, four, and five.  The calibration models for the ID-test data are used to calibrate the OOD uncertainty estimates; the OOD residuals are not used in order to prevent data leakage in the active learning process. In keeping with the higher average residual value, the LL-ensemble uncertainty estimates are larger than the random-ensemble uncertainty estimates, resulting in OOD error bars that better overlap with the parity line in (h, i, j) compared to the random-ensemble error bars (r, s, t). However, the larger error bars in (c, d, e) are unnecessary for the ID-test data, where smaller uncertainty values such as those in (m, n, o) are sufficient to overlap with the parity line. The large uncertainty of OOD samples relative to ID samples motivates the high-uncertainty acquisition policy in the active learning study. 

\begin{figure}[!h]
    \centering
    \includegraphics[width=\linewidth]{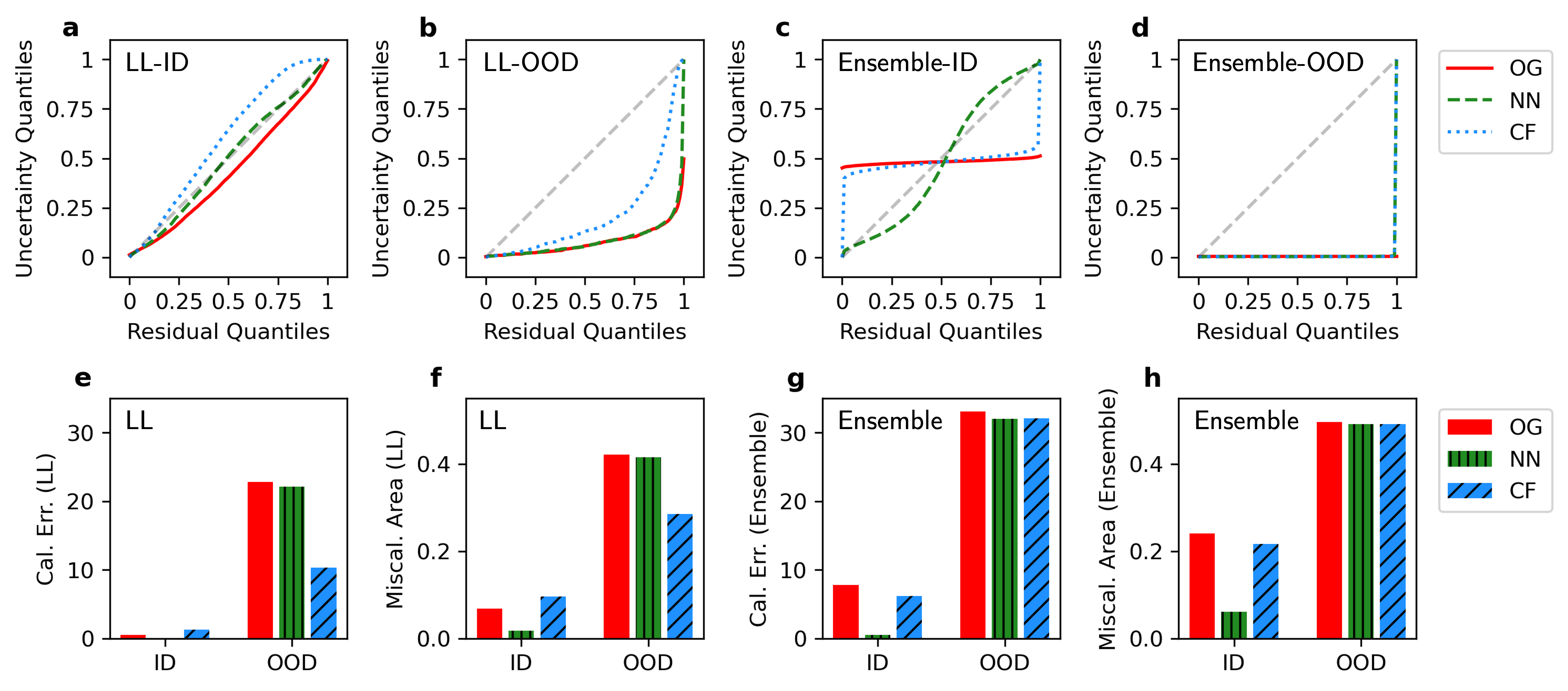}
    \caption{Model trained omitting F-containing compounds to predict formation energy $E_f$, yielding a comparison of neural network (NN) and calibration factor (CF) calibration methods to the original uncertainty (OG) distributions. (a) ID-test uncertainties from LL-ensemble. (b) OOD uncertainties from LL-ensemble. (c) ID uncertainties from random ensemble. (d) OOD uncertainties from random ensemble. (e) Calibration errors from LL uncertainty estimates in (a) and (b). (f) Calibration errors from LL uncertainty estimates in (a) and (b). (g) Calibration errors from random ensemble uncertainty estimates in (c) and (d). (h) Miscalibration area from random ensemble uncertainty estimates in (c) and (d).}
    \label{fig:calibration_comparison}
\end{figure}

\begin{figure}[!h]
    \centering
    \includegraphics[width=\linewidth]{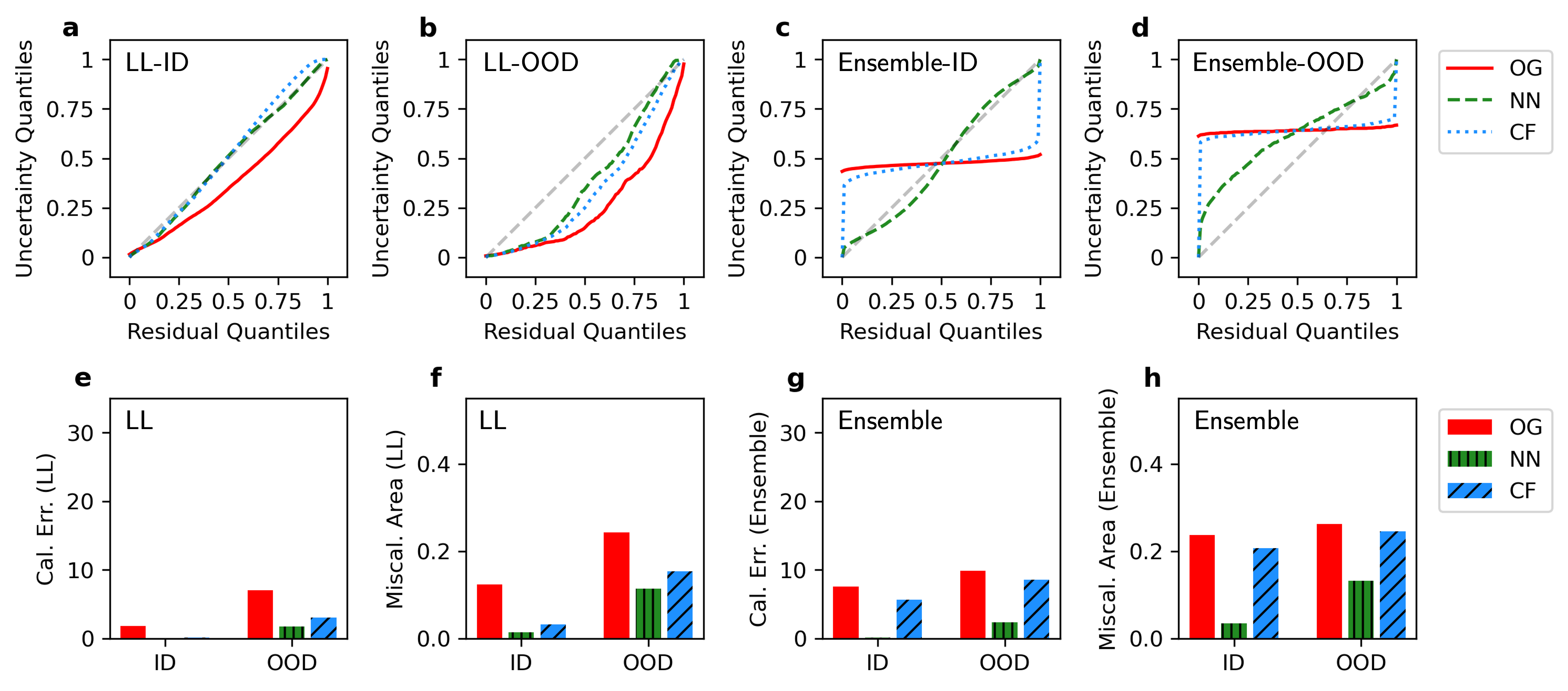}
    \caption{Model trained omitting Fe-containing compounds to predict formation energy $E_f$, yielding a comparison of neural network (NN) and calibration factor (CF) calibration methods to the original uncertainty (OG) distributions. (a) ID-test uncertainties from LL-ensemble. (b) OOD uncertainties from LL-ensemble. (c) OOD uncertainties from random ensemble. (d) OOD uncertainties from random ensemble. (e) Calibration errors from LL uncertainty estimates in (a) and (b). (f) Calibration errors from LL uncertainty estimates in (a) and (b). (g) Calibration errors from random ensemble uncertainty estimates in (c) and (d). (h) Miscalibration area from random ensemble uncertainty estimates in (c) and (d).}
    \label{fig:calibration_eform_Fe_alignn}
\end{figure}

The differences between the ID-test and OOD uncertainty estimates from the LL-ensembles and random-ensembles, and the impact of calibration methods can be quantified further using quantile-quantile ($Q-Q$) plots. As shown in Figure~\ref{fig:calibration_comparison}(a) for the ID-test uncertainty distributions visualized in Figure~\ref{fig:eform_F_uncert_hist_parity}(b), 
the cumulative distribution function (CDF) of the uncertainties for ID-test or OOD data is taken as the dependent variable (labeled uncertainty quantiles)
and plotted against the CDF of the ID-test residuals (labeled residual quantiles).

Perfect agreement between the two distributions is represented by the straight dashed line; any deviation from this line represents a miscalibration. The miscalibration between the two distributions is quantified using the calibration error and the miscalibration area~\cite{tran2020methods}. The analysis is repeated for the OOD LL-ensemble distributions in Figure~\ref{fig:calibration_comparison}(b), the ID-test random-ensemble distributions in (c), and the OOD random-ensemble distributions in (d). The comparison of the LL-ensemble ID-test and OOD metrics are shown in Figure~\ref{fig:calibration_comparison}(e, f), and the comparison of the random-ensemble ID-test and OOD metrics are shown in Figure~\ref{fig:calibration_comparison}(g, h). In (a-d), closer to the dashed line represents a better calibrated distribution, below the line suggests over confidence and above the line suggests under confidence. In (e-h), closer to zero represents a better calibration. In Figure~\ref{fig:calibration_comparison}, the uncalibrated LL-ensemble uncertainties are in better alignment with both the ID-test and OOD data residual distributions than the random-ensemble uncalibrated uncertainties. The NN calibration method out-performs the CF calibration methods for the ID-test data, reducing the miscalibration to near zero.  However, the CF calibration method generalizes slightly better to the OOD data uncertainties. 

From the LL analysis in Figure~\ref{fig:loss_landscapes}(g) and (l), we can consider the difference between the ID-test and OOD data distributions for the formation energy task with fluorine-containing compounds to be significant since the loss landscapes are distinct. Repeating the same uncertainty analysis for a study where the ID-test and OOD loss landscapes are more similar, such as the formation energy prediction task when iron (Fe) is omitted. The loss landscape similarity is shown in Figure~\ref{fig:loss_landscapes}(h) and (m), and the results of applying the uncertainty quantification calibration demonstrated in Figure~\ref{fig:eform_F_uncert_hist_parity} to the $E_f$-Omit Fe study are shown in Figure~\ref{fig:calibration_eform_Fe_alignn}. 

These trends only partially continue for the remaining prediction tasks and models shown in the Supplement. For example, the uncalibrated ALIGNN random-ensemble uncertainties outperform the uncalibrated LL-ensemble uncertainties for the bandgap $E_g$ prediction tasks for both the omit-F and omit-Fe cases, and the CF-calibration method collapsed for the bandgap $E_g$ prediction tasks because the distribution of ID-test residuals was non-Gaussian. The NN-calibration method consistently outperforms the CF-calibration method on the ID-test data, and even when the CF-calibration method failed on the ID-test data it still improved the calibration of the OOD-test data over the uncalibrated estimate. For XGB and RF models, intrinsic model uncertainties~\cite{brophy2022ibug, meinshausen2006quantile} are used in-place of the LL-ensemble. A comparison of the intrinsic model uncertainties to the query-by-committee (QBC) uncertainty used during active training shows that the uncalibrated intrinsic model uncertainty has a lower calibration error compared to the uncalibrated QBC uncertainty, and that the NN uncertainty calibration method generally outperforms the CF uncertainty calibration method on ID-test and OOD data. 

\subsection*{Implications for active learning tasks}

Now that the uncertainty estimates are prepared, their usefulness can be evaluated by an active learning experiment that selects samples with maximum uncertainty as its acquisition policy. If the relative uncertainty estimate for each sample is a meaningful representation of the information required to generalize to the OOD-test dataset, selecting samples with higher uncertainty should help the model improve more quickly than random selection. For the active learning task, 20~\% of the OOD data was reserved as OOD-test to validate the active learning success. For each active learning experiment, a model pretrained on the ID-train data is re-trained using the original ID-train data partition with $X~\%$ OOD-train data added ($X\in [5, 100]$). The active learning experiment was repeated five times with five different random seeds.

\begin{figure}[!h]
    \centering

    \begin{subfigure}[b]{\textwidth}
            \includegraphics[width=\linewidth]{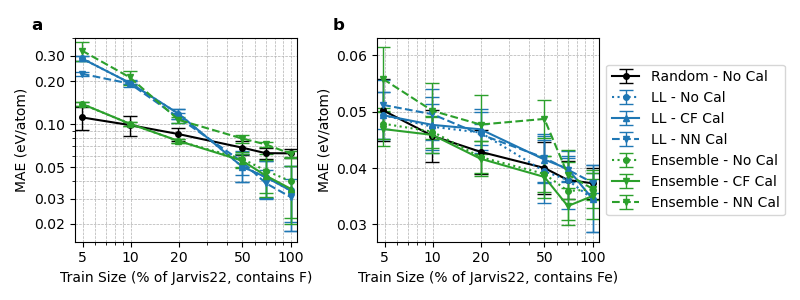}
    \end{subfigure}
    \caption{Active learning implemented for formation energy task for uncertainty maximization acquisition policy given different uncertainty estimates; lower MAE is better. (a) The OOD-train pool consists of fluorine-containing compounds. (b) The OOD-train pool consists of iron-containing compounds. For the fluorine omission study (a), random-ensemble uncertainty estimates generally out perform LL methods. For the iron omission study (b), uncertainty source and calibration method of the uncertainty does not significantly impact the the dataset.}
    \label{fig:active_learning_eform}
\end{figure}

In Figure \ref{fig:active_learning_eform}, active learning results for the formation energy task are presented for each of the uncertainty sources and calibration methods previously reported; results for the bandgap task are in the Supplement. The OOD-train dataset consists of fluorine-containing compounds in Figure \ref{fig:active_learning_eform}(a), and iron-containing compounds in Figure \ref{fig:active_learning_eform}(b). The mean prediction error of all trials is presented, and the standard deviation of the mean prediction error is presented as error bars. The uncertainty maximization acquisition policy is compared to random sampling as a baseline. 

In Figure~\ref{fig:active_learning_eform}, the overlapping error bars indicate there is no best single method for acquiring OOD samples, and calibrating the uncertainty may not improve the original random ensemble uncertainty estimate. For small OOD-train samples and the fluorine-containing OOD-train dataset, random sampling out performs or performs comparably with uncertainty-maximization methods regardless of calibration method until 20~\% of the OOD-train data has been implemented. The calibration factor (CF) corrected uncertainty results closely follow the uncalibrated uncertainty methods; this confirms that the linear scaling is not expected to affect the order with which samples are introduced during active learning. When the OOD-training pool is fluorine-containing elements, the LL uncertainties in Figure~\ref{fig:active_learning_eform}(a) do not outperform the random sampling acquisition policy regardless of whether they are calibrated or uncalibrated until $\approx50~\%$ of the OOD training pool has been added to the training dataset. Similarly, for Figure~\ref{fig:active_learning_eform}(b), the uncalibrated and CF-calibrated random ensemble uncertainties consistently outperform the random acquisition policy, but only within the margin of error due to random initialization.  The LL uncertainties again perform comparably or worse than the random acquisition regardless of uncertainty calibration method. This indicates that although the calibration methods may significantly affect the distribution shape and scale of the estimated uncertainties for the OOD data, the information represented by the calibrated uncertainty estimates did not help the model improve.

Figure~\ref{fig:active_learning_eform}(a) represents a study where the OOD data is not well represented by the original weight space of the model as seen in the loss landscapes and uncertainty distributions, while in Figure~\ref{fig:active_learning_eform}(b) the OOD data partition could be considered as approximately ID. The loss of Figure~\ref{fig:active_learning_eform}(a) changes by an order of magnitude from 0.3 eV/atom to 0.06 eV/atom, while the loss of Figure~\ref{fig:active_learning_eform}(b) stays in the range of 0.06 eV/atom to 0.03 eV/atom. This indicates that while uncertainty calibration may not have improved model performance, it also did not harm the model's ability to acquire new samples if those samples were approximately ID with respect to the original training distribution.

\begin{figure}
    \centering
    \includegraphics[width=\linewidth]{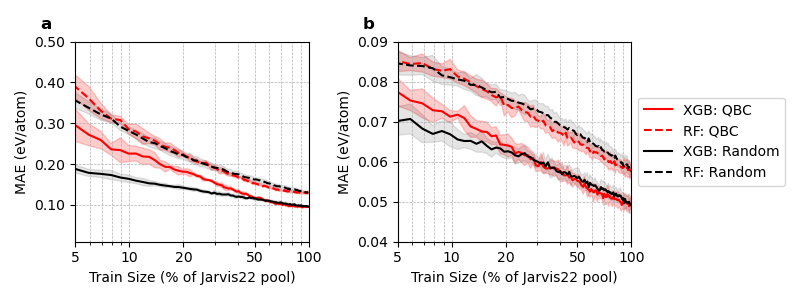}
    \caption{Active learning implemented for formation energy task for XGB and RF models. (a) The OOD-Train Jarvis-22 pool consists of fluorine-containing compounds. (b) The OOD-Train Jarvis-22 pool consists of iron-containing compounds. In both cases, random acquisition performs as well or better than the uncertainty maximization policy, indicating that the uncertainty estimate does not approximate the missing information necessary for the model to generalize to the remaining OOD data.}
    \label{fig:active_xgb_rf_eform}
\end{figure}

The study from Figure~\ref{fig:active_learning_eform} is repeated in Figure~\ref{fig:active_xgb_rf_eform}, with the same data used to train RF and XGB models. Here, the uncertainty for each sample is the disagreement between models in a committee of models, and the acquisition policy again maximizes uncertainty. A similar pattern to Figure~\ref{fig:active_learning_eform} appears, where the random sampling performs as well or better than the uncertainty maximization policy for both the RF and the XGB models. 

The results for the bandgap included in Supplement Figure~\ref{fig:active_bgap} indicate a similar trend. When the OOD-test dataset is fluorine-containing compounds, the data is strongly OOD as shown in Figure~\ref{fig:loss_landscapes} and no calibrated or uncalibrated uncertainty estimation method in this work outperformed random sampling. However, for the bandgap test when the OOD-train dataset is iron-containing compounds, the calibrated ensemble uncertainties, uncalibrated ensemble uncertainties, and the calibration factor (CF) method for the LL uncertainties outperform random sampling on average -- the error bars continue to overlap. This is also seen in Figure~\ref{fig:active_xgb_rf_bgap}(b) for the XGB and RF study using the same data; the maximum uncertainty acquisition policy outperforms the random sampling after 20~\% of the OOD-training data has been added to the larger training dataset.

% \clearpage
\section*{Discussion}
\label{sec:discussion}

The challenges associated with using models from different data distributions have been previously reported~\cite{rahaman2021uncertainty, Ashukha2020Pitfalls}. Consider a hypothetical case where a second model has been trained exclusively with samples that are OOD with respect to a previous model trained on ID data samples; averaging the predictions of these models is expected to result in prediction uncertainties that are too large~\cite{rahaman2021uncertainty}. When sampling the loss landscape to create the set of seed models for the LL-ensemble, a higher loss threshold allows models farther from the ID loss landscape minima and models closer to the OOD loss landscape minima to be averaged together; there is currently no heuristic that identifies an optimal threshold. As a result, more models outside of the original model's training distribution are added to the $\mathcal{N}\left(\mu_\theta, \sigma_\theta\right)$ distribution. At the other extreme, uncertainty estimates created by a narrow distribution of model ensembles tend to be over confident~\cite{Ashukha2020Pitfalls} with error bars that are too small. This is seen in Figure~\ref{fig:eform_F_uncert_hist_parity} when error bars from the random ensemble of models all trained to a similar loss value are too small to cross the parity line. Including the Normal distribution in the posterior when sampling the loss landscape estimates alleviates some of this issue by relaxing parameter correlations within the model and allowing more diversity in the ensemble. Finally, using specific Hessian-eigenvectors to construct the loss landscape, such as the largest-eigenvalued eigenvectors to probe the loss landscape minima or largest-eigenvalued and smallest/most-minimum eigenvalued eigenvectors to probe any loss landscape saddle point, provides a level of geometric control over the topology of the loss landscape. With the geometric manipulation of the loss landscape, the introduction of loss-thresholded sampling from the loss landscape and a relaxation of model parameter correlations, loss landscapes provide accessible and explainable control of the posterior for the ensemble, resulting in error bars which are somewhat tunable. The ensemble of models used for prediction can be intentionally constructed to have certain properties (i.e., prioritizing a subset of samples or one objective out of many in a multiobjective task), so that uncertainty estimates will therefore reflect specific properties of the data.

Uncertainty calibration maps uncertainty estimates to a known distribution. Empirical reference distributions leverage the training data because this is the data used to optimize the model; another common practice chooses an arbitrary distribution, e.g. $\mathcal{N}(0, 1)$, as the reference distribution. In this work, we first confirmed that when the test data is ID with respect to the training data and the statistical model is taken from the ID residuals, the calibration error of the original uncertainty is usually low. Notable exceptions are the ALIGNN LL-ensemble uncertainty estimates for the bandgap prediction task; these are shown in the supplement. The uncertainty calibration methods applied to the ID data uncertainties are expected to have minimal impact because there is minimal calibration error, but significant changes in the median uncertainty value and the shape of the uncertainty distribution were observed in Figure~\ref{fig:eform_F_uncert_hist_parity} and in the remaining supplementary figures. In some cases, the calibration factors method of uncertainty calibration had the effect of increasing rather than decreasing the miscalibration for the ID uncertainty estimates, especially when the residuals did not have a Normal distribution. The use of the neural network to calibrate uncertainties did not suffer from this increase in miscalibration error. This suggests that the assumptions underlying the calibration method are as important to the quality of the uncertainty calibration as the choice of reference distribution. 

The use of the ID uncertainty calibration model on the OOD uncertainty estimates propagates any assumptions regarding the ID uncertainty distribution, including the choice of calibration method and choice of calibration reference distribution. A reduction in OOD uncertainty miscalibration error seems loosely correlated with a reduction in ID uncertainty miscalibration. This is demonstrated by cases where the OOD miscalibration increases while the ID miscalibration decreases, and by cases where the OOD miscalibration decreases while the ID miscalibration increases as in supplementary Figure~\ref{fig:calibration_eform_Fe_rf} for the CF calibration method. The OOD uncertainty miscalibration can be expected to improve with the ID calibration model when the OOD uncertainty distribution correlates with ID residual distribution. A lack of joint improvement in ID and OOD miscalibration therefore quantifies how the ID and OOD distributions differ. 

In the active learning task, the various uncertainty estimates were assumed to correlate with the epistemic uncertainty of the model; the model is missing the information needed to predict the true value for the sample with confidence. The use of calibrated uncertainties was found to have little effect or negative effect on a model's performance compared to random sampling. In cases where model performance degraded with a calibrated uncertainty compared to the uncalibrated uncertainty, the calibration model effectively removed information necessary for the model to improve by deprioritizing the more-uncertain samples in favor of statistically more likely samples relative to the model. This provides insight into a situation where a sample can be statistically more-likely with respect to the dataset distribution, and statistically less-likely with respect to the distribution learned by the model~\cite{zhou2023samples, arpit2017closer}. As a result, a poor quality uncertainty estimate cannot be corrected using calibration methods. When data is expected to be OOD, prediction variance produced by an ensemble may be unable to capture the information missing in the predictive model, and methods such as counterfactual explanations~\cite{verma2024counterfactual} may be more useful.

Finally, the effect of the different uncertainty calibration methods on the acquisition policy cannot be fully understood from the prediction space alone. In the Supplement, histograms of target distribution changes as OOD-train data is added show minimal difference between uncertainty sources and calibration methods, and do not explain cases where some uncertainty estimates or calibration methods outperform other methods. The LL-ensemble based uncertainties were marginally less efficient in the active learning campaign than the random-ensemble uncertainty estimates; the target distributions do not predict this outcome. This suggests that uncertainty estimates from the input feature space may ultimately prove more meaningful.

% \clearpage
\section*{Methods}
\label{sec:methods}

\subsubsection*{Estimated Solubility (ESOL) dataset}
\label{sssec:esol_data}

The Estimated Solubility (ESOL) dataset \cite{delaney2004esol} from the MoleculeNet benchmark \cite{wu2018moleculenet} contains experimental water solubility values for 1,128 compounds. Chemical structures in the ESOL dataset were converted from SMILES strings \cite{delaney2004esol} to a Morgan fingerprint \cite{morgan1965generation} that encodes substructures present in the molecule connectivity as a fixed length bit string at the cost of chemical information \cite{tayyebi2023prediction}. The molecular structures at the periphery of the molecule significantly impact its solubility, which in turn impacts its bioavailability \cite{capecchi2020one} and therefore motivates our choice of this representation for this work. Principal Component Analysis (PCA) \cite{greenacre2022principal} was then applied for dimensionality reduction, followed by Density-Based Spatial Clustering of Applications with Noise (DBSCAN) \cite{ester1996density, schubert2017dbscan} to identify clusters of structurally similar molecules. Two clusters are set: the data points in the core cluster are in-distribution and the rest of the data points are defined as out-of-distribution. The resulting clusters highlight molecular subgroups with potentially shared solubility characteristics, providing insights into structure-property relationships. The clusters are shown in Supplement Figure \ref{fig:dbscan_cluster}.

\subsubsection*{JARVIS-22 3D DFT Dataset}

Two material properties are selected for this study: the optb88vdw electronic bandgap $E_g$ and formation energy $E_f$. These properties are meaningful for a variety of tasks, and are a primary outcome of first-principles calculation methods \cite{choudhary2020joint}. The preparation of the data and training of the ALIGNN models to predict the bandgap and formation energy follows the method presented in \cite{li2025probing}. An element $X \in $ [F, Fe] is selected to partition the dataset into an in-distribution (ID) dataset that omits any chemistry containing element $X$, and an out-of-distribution dataset that contains chemistries which include element $X$. 

For the QBC:XGB and QBC:RF models, materials in JARVIS-22 with formation energy larger than 5 eV/atom were removed. Features were generated by Voronoi tessellation featurizer~\cite{ward2017including} in Matminer~\cite{ward2018matminer}, giving 273 compositional and structural features for each material. Some compounds failed tessellation and were excluded. For the ALIGNN models, data preparation strictly followed~\cite{li2025probing}.

\subsection*{Models}

The solubility task implemented a three layer multilayer perceptron (MLP) architecture in PyTorch 2.7~\cite{paszke2019pytorch} with drop out and ReLU activation between the MLP layers. The model was trained for 100 epochs with a batch size of 32 using mean squared error loss and an Adam optimizer with weight decay of 1e-5. Hyper parameters, including learning rate, hidden dimension, the number of hidden layers, and drop out rate were optimized using Bayesian optimization; this is discussed further in the supplement. 

ALIGNN architectures \cite{choudhary2021atomistic} were used for the loss landscape studies of formation energy and bandgap regression tasks. The ALIGNN models were prepared and trained following the method presented in~\cite{li2025probing}. Each model consists of feature embedding layers followed by two ALIGNN layers, two graph convolution layers, and a multiperceptron layer with size 256. The models were trained for 25 epochs using a batch size of 128 and the adamw optimizer. The full model configuration and training specifications are available with the code.

Random Forest (RF)~\cite{breiman2001random} and Extreme Gradient Boosting (XGB)~\cite{chen2016xgboost}. RF models were implemented in scikit-learn 0.24.2~\cite{pedregosa2011scikit} with 100 trees and \lstinline{max_features=0.3}. XGB models were implemented in xgboost 2.1.4~\cite{chen2016xgboost} with 4 parallel trees, 1000 estimators per tree, learning rate 0.1, L1 regularization 0.01, L2 regularization 0.1, histogram grow method, and subsampling ratios 0.85 (rows), 0.3 (columns per tree), 0.5 (columns per level)~\cite{li2023exploiting}. Hyperparameters excepting the random seed were fixed due to negligible impact on performance.

\subsection*{Loss Landscape Study}
\label{ssec:loss_landscape}

For model weights $\theta$, the loss landscape is calculated by interpolating in the weight space and calculating the population loss for each interpolated point. The interpolation directions are chosen to be the largest positive magnitude Hessian eigenvector $\eta$ and the second largest minimum Hessian eigenvector $\delta$ so that the population loss is 
\begin{equation}
    \mathcal{L}(\theta + \alpha \eta + \beta \delta) = \frac{1}{N} \sum_{i=1}^N \ell \left(x_i, y_i; \theta + \alpha \eta + \beta \delta \right) .
    \label{eq:2D_loss_landscape}
\end{equation}
Accordingly, the sample-target pair $(x_i, y_i)$ may be drawn from either the in-distribution or out-of-distribution datasets, resulting in distinct loss landscapes for the in-distribution and out-of-distribution datasets. The constants $\alpha, \beta$ represent the interpolation step in the weight space; $\alpha = \beta = 50$. The model weights $\theta$ can be centered in the loss landscape using a shift of coordinates. The scale for all loss landscapes $-\eta/2 < \alpha \leq \eta/2$ and $-\delta/2 < \beta \leq \delta/2$ for eigenvectors $\eta, \delta$. 

\subsubsection*{Estimation of Uncertainties}
\label{ssec:calc_of_err}

To create a calibrated uncertainty model, a set of uncertainties is divided into test and train partitions to prevent overfitting to the distribution and ensure generalizability. In this work, the train partition is chosen to be the ID test data, and the test partition is the OOD data.

Prediction variances from the loss landscapes were obtained using an adaptation of stochastic weight averaging with a Gaussian assumption \cite{maddox2019simple} to create a diagonal covariance matrix; the adaptation uses the Hessian loss landscape of the trained model rather than model samples along the stochastic weight averaging trajectory, following the method presented in \cite{ravishankar2022stochastic}. During the calculation of each loss landscape and for a given performance threshold $\varepsilon$, any set of model weights $\theta_ij = \theta + \alpha_i \eta + \beta_j \delta$ with population loss $\mathcal{L}(\theta_{ij}) <\varepsilon$ were reserved as part of a set of $n$ models $\Theta$ with equivalent performance, such that $\Theta = [\theta_1, \theta_2, ..., \theta_n]$. A Normal distribution of models $\mathcal{N}(\mu_\Theta, \sigma_\Theta)$ may be defined using 
\begin{equation}
    \mu_\Theta = \frac{1}{n}\sum_i^n \theta_i\ \text{for}\ \theta_i \in \Theta
    \label{eq:mean_theta}
\end{equation}
and the standard deviation of each parameter in the distribution may be calculated from
\begin{equation}
    \sigma_\Theta = \sqrt{\frac{1}{n}\sum_i^n \left(\theta_i - \mu_\Theta \right)^2}\  \text{for}\ \theta_i \in \Theta .
    \label{eq:sigma_theta}
\end{equation}
This approach requires a second pass through the loss landscape, but was found to be sufficiently numerically stable for our purposes. Through the use of in-place operations, this method requires three copies of the model architecture to be maintained in memory. 

After calculation, the $\mathcal{N}(\mu_\Theta, \sigma_\Theta)$ distribution is sampled repeatedly to generate additional models with similar performance to the original model, but with perturbed weights. A benefit of sampling models from $\mathcal{N}(\mu_\Theta, \sigma_\Theta)$ is that the uncertainty calculation is no longer restricted to models drawn from the loss landscape calculation; models statistically similar to the reference model weights may also be included in uncertainty estimations since the effect of the standard deviation in the $\mathcal{N}(\mu_\Theta, \sigma_\Theta)$ is to perturb model weights in directions unaligned with the original eigenvectors used to orient the loss landscape plane. In this work, the performance threshold $\varepsilon$ is 0.1. The distribution $\mathcal{N}(\mu_\Theta, \sigma_\Theta)$ is sampled 50 times, and used to create 50 sets of predictions for the in-distribution test set and the out-of-distribution test set. 

The intrinsic uncertainty of the each XGB and RF model for each prediction was determined using the method of~\cite{brophy2022ibug} for XGB models and~\cite{meinshausen2006quantile} for RF models respectively.  The query-by-committee (QBC) uncertainty used during active training is taken as the prediction disagreement between an ensemble of an RF model and an XGB model after the method in~\cite{li2023exploiting}. 

\subsubsection*{Uncertainty Calibration}

Following the calibration factor method in~\cite{palmer2022calibration} and adopting the notation, a linear relationship between the calibrated and uncalibrated uncertainties $\hat{\sigma} = a\hat{\sigma}_{uc} + b$ is assumed, so that optimizing
\begin{equation}
    a, b = \text{argmin}_{a', b'} \sum_{x, y \in D_{cv}} \text{ln}\left(a' \hat{\sigma}_{uc}\left(x\right) + b' \right)^2 + R\left(x\right)^2/\left(a'\hat{\sigma}_{uc}\left(x\right) + b' \right)^2
\end{equation}
for cross-validation data $D_{CV}$ and the residual $R(x)$ of sample $x$ yields calibration factors $a, b$. The equation is optimized using the ID-test residuals and uncertainties only, then the calibration factors are used to scale the OOD uncertainty estimates. Unique calibration factors were obtained for each task, model, and uncertainty estimate; the table of calibration factors obtained is reported in the Supplement. The software implementation for the calibration factor method is adapted from the open source repository associated with~\cite{palmer2022calibration}. 

The neural network calibration method proposed in this work approximates the method of histogram-based uncertainty calibration~\cite{kumar2019verified} using a differentiable soft-max function to ``bin'' values and create an approximation of the reliability plots shown in Figures~\ref{fig:calibration_comparison} and~\ref{fig:calibration_eform_Fe_alignn}. The loss function minimizes the distance between the $Q-Q$ line and the parity line determined by the ID-test residuals. A fully connected neural network with three layers [2, 32, 1] and ReLU activation functions was trained to predict a calibrated uncertainty value given an input prediction value and associated estimated uncertainty. The code for the loss function and model training losses are provided in the supplement. We do not assert that this approach is well-principled, but propose that it is effective.

% \clearpage
% \input{sec/6_conclusion}
% % \clearpage

\subsubsection*{Data \& Code Availability} 
The ESOL dataset is available through the MoleculeNet benchmark \cite{wu2018moleculenet}. The JARVIS 3D DFT dataset used in this work is available from Zenodo \cite{li_2024_12763938}. Code will be released on GitHub upon acceptance.

\subsubsection*{Acknowledgements}
A.S.D. acknowledges supports from the University of Toronto’s Eric and Wendy Schmidt AI in Science Post-doctoral Fellowship, a program of Schmidt Sciences. Y.H. is supported by a summer research studentship from the University of Toronto’s Acceleration Consortium initiative, which receives funding from the Canada First Research Excellence Fund (CFREF).

\subsubsection*{Author Contributions}
A.S.D, K.L., J.H.-S. conceived and designed the project. K.L., A.S.D., H.W., Y.H. trained ML models and/or performed Hessian loss landscape calculations. A.S.D., E.H. analyzed the results. A.S.D. drafted the manuscript. J.H.-S supervised the project. A.S.D., K.L., B.D., and J.H.-S. discussed the results. A.S.D., K.L., B.D., and J.H.-S. reviewed and edited the manuscript. All authors contributed to the manuscript preparation.

\subsubsection*{Competing Interests}
All authors declare no financial or non-financial competing interests. 
Certain commercial products or company names are identified here to describe our study adequately. Such identification is not intended to imply recommendation or endorsement by the National Institute of Standards and Technology, nor is it intended to imply that the products or names identified are necessarily the best available for the purpose.

% \clearpage
{
    \small
    \bibliography{main}
}
\clearpage
\begin{appendices}
\beginsupplement
\clearpage

\section*{Supplementary Information}

\subsection*{Dataset Details}
\label{ssec:data_deets}
\begin{table}[h]
    \centering
    \caption{ESOL Dataset Partitions}
    \begin{tabular}{cc}
         \textbf{PARTITION} &  \textbf{\# SAMPLES}\\
         \hline
         Training & 867\\
         Training Validation & 97 \\
         In-distribution Test & 108 \\
         Out-of-distribution Test & 56
    \end{tabular}
    \label{tab:esol_data_splits}
\end{table}

\begin{figure}[!h]
    \centering
    \includegraphics[width=0.5\linewidth]{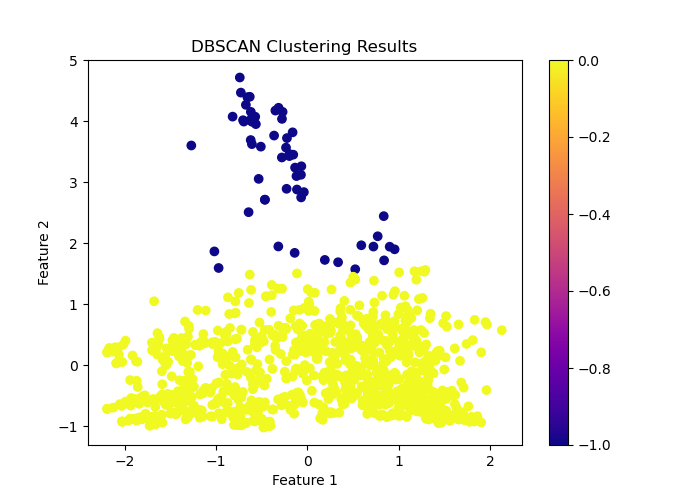}
    \caption{DBSCAN Clustering results to partition the ESOL dataset into in-distribution (blue) and out-of-distribution (orange) samples. DBSCAN hyperparameters include $\varepsilon = 0.5$ and }
    \label{fig:dbscan_cluster}
\end{figure}

The results from clustering the data using DBSCAN after PCA decomposition are shown in Figure \ref{fig:dbscan_cluster}. The standard implementations of DBSCAN and PCA from Scikit-Learn \cite{scikit-learn} v. 1.6.1; hyperparameters were left to default unless mentioned.

\begin{table}[h]
    \centering
    \caption{Dataset Partitions for Loss Landscape Study}
    \begin{tabular}{cccc}
        \textbf{OMITTED ELEMENT} & \textbf{\# TRAIN} & \textbf{ \# ID TEST} & \textbf{\# OOD TEST} \\
        \hline
        F & 56438 & 14110 & 1087 \\
        Fe & 57520 & 14380 & 817\\
    \end{tabular}
\end{table}

\begin{table}[h]
    \centering
    \caption{Dataset Partitions for Active Learning Study}
    \begin{tabular}{ccccc}
        \textbf{OMITTED ELEMENT} & \textbf{\# ID TRAIN} & \textbf{\# OOD TRAIN} & \textbf{ \# ID TEST} & \textbf{\# OOD TEST} \\
        \hline
        F & 56438 & 14110 & 7148 & 217\\
        Fe & 57520 & 653 & 14380 & 163\\
    \end{tabular}
\end{table}

\subsection*{Model Training Details}
\label{ssec:model_train_deets}

\subsubsection*{Bayesian Optimization Process of Neural Network for Solubility Study}

Bayesian Optimization (BO) is a global optimization technique that efficiently searches for optimal hyperparameters by constructing a probabilistic model of the objective function. Unlike grid or random search, BO strategically selects hyperparameter values by balancing exploration (sampling uncertain regions) and exploitation (sampling promising regions). This is achieved using a Gaussian Process (GP) to model the objective function, coupled with an acquisition function (e.g., Expected Improvement or Upper Confidence Bound) to determine the next sampling point.

In this study, BO is employed to optimize hyperparameters of a PyTorch-based deep learning model trained on the ESOL dataset for aqueous solubility prediction. The optimization objective is to minimize the R\textsuperscript{2} on the validation dataset. The process iteratively evaluates different hyperparameter configurations and converges to an optimal set through multiple iterations.

\subsubsection*{Optimization Process Execution}
The Bayesian Optimization framework is implemented using the bayes\_opt Python library. The process includes:

\begin{itemize}
    \item Random Sampling (Initialization): The optimizer evaluates 5 randomly selected hyperparameter sets to initialize the search space.
    \item Iterative Optimization: Over 15 additional iterations, the optimizer selects hyperparameters using the acquisition function to maximize improvement.
    \item Model Training \& Evaluation: For each sampled hyperparameter set, the model is trained for 50 epochs and evaluated using MSE.
    \item Convergence to Optimal Hyperparameters: The best-performing hyperparameter set is selected and used for the final model.
\end{itemize}

\begin{figure}[!h]
    \centering
    \includegraphics[width=0.75\linewidth]{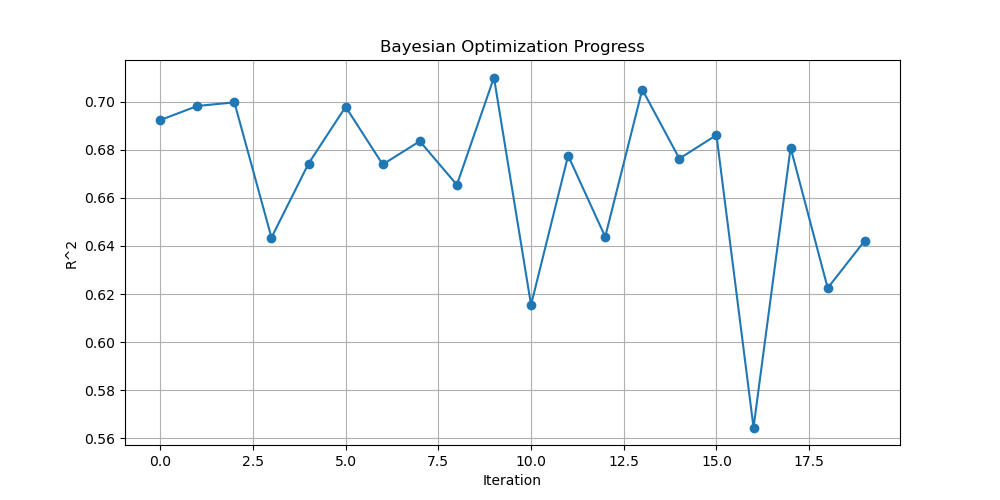}
    \caption{Bayesian Optimization Process}
    \label{fig:nn_bayes_opt}
\end{figure}

\subsection*{Hessian Vector Product Estimation}
\label{ssec:hvp_est}
Consider a model with weights $\boldsymbol{\theta}$ and population loss $\mathcal{L}$. The Hessian of the loss is then \cite{singh2021analytic}
\begin{equation}
    \boldsymbol{H}_{\theta} = \frac{\partial^2 \mathcal{L}}{\partial \boldsymbol{\theta} \partial \boldsymbol{\theta}} .
\end{equation}
The practical implementation of this calculation requires implementing backpropagation twice over the model's parameters
%\cite{}
. For models with many parameters, this calculation quickly becomes intractable 
% \cite{}
. Therefore, the Hessian eigenvectors are commonly approximated using the Hessian vector product method, which avoids calculating the full Hessian directly. Given a vector $v \in {\rm I\!R}^N$, then the Hessian vector product is calculated
\begin{equation}
    \nabla_\theta \left[ \left( \nabla_\theta \mathcal{L} \right)^{T} v\right] = \left(\nabla_\theta \nabla_\theta \mathcal{L} \right)v + \left(\nabla_\theta \mathcal{L} \right)^T \nabla_\theta v = H_\theta v
\end{equation}
following the method presented in \cite{bottcher2024visualizing}. The largest-magnitude eigenvector $\eta$ is calculated first, followed by the second-largest magnitude eigenvector of opposite sign, $\delta$. These eigenvectors are then used as directions in the loss landscape. 

In this work, 500 training samples were randomly selected from the in-distribution test set were used to estimate the population loss $\mathcal{L}$ in the Hessian vector product calculation.

\clearpage
\subsection*{Linear-Scale Loss Landscapes}
\label{ssec:lin_scale_ll}

\begin{figure}[!h]
    \centering
    \includegraphics[width=\linewidth]{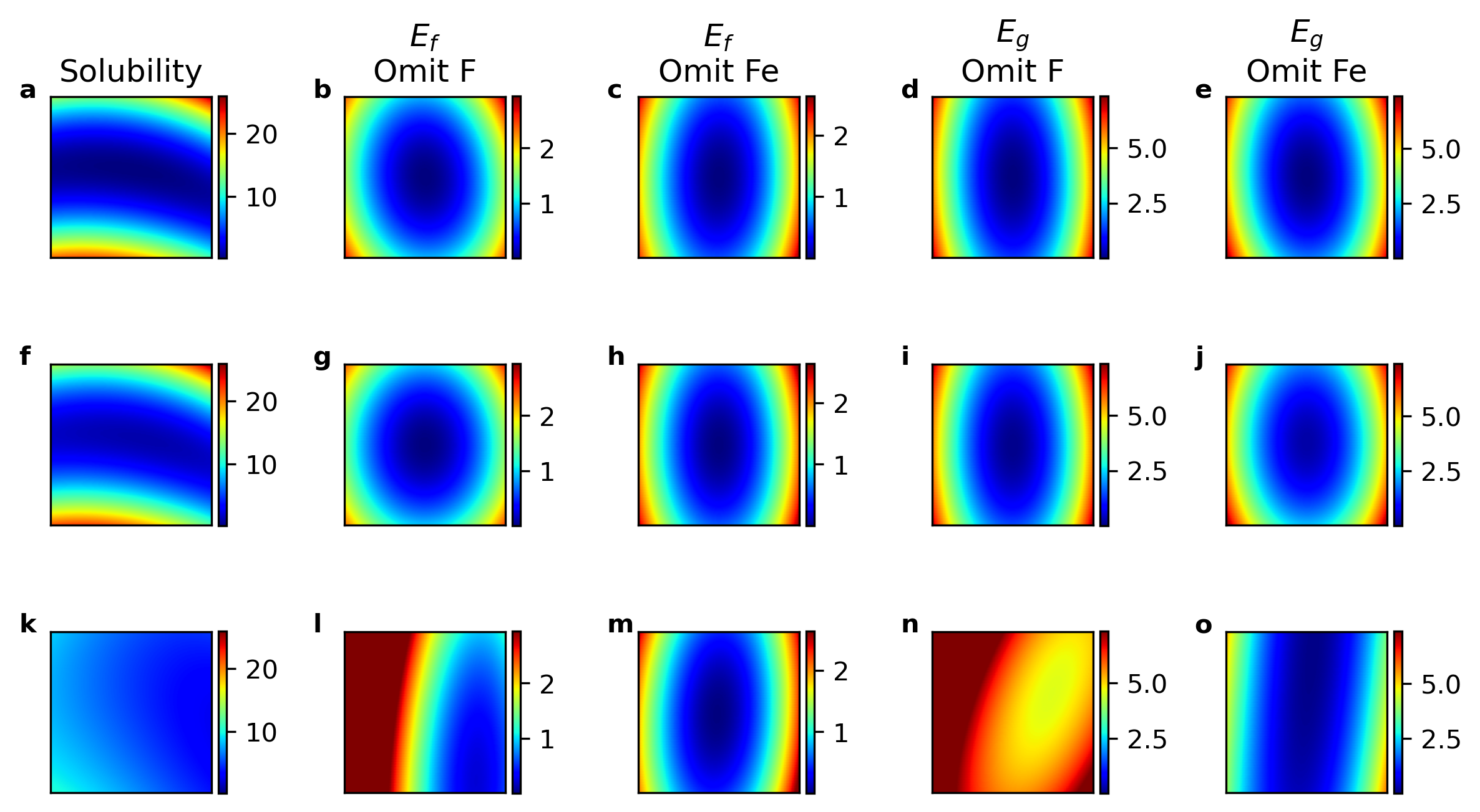}
    \caption{Linear scale loss landscapes. (a-e) Training data. (f-j) ID-test data. (k-o) OOD-test data.}
    \label{fig:linear_ll}
\end{figure}

\clearpage
\subsection*{Uncertainty Calibration}

In the following figures: (a) The calibration of the ID-test data uncertainties from the LL-ensemble estimate. (b) The calibration of the OOD data uncertainties from the LL-ensemble estimate. (c) The calibration of the ID-test data uncertainties from the random-ensemble estimate. (d) The calibration of the OOD data uncertainties from the random-ensemble estimate. (e) The calibration error from (a) and (b). (f) The miscalibration area from (a) and (b). (g) The calibration error from (c) and (d). (h) The miscalibration area from (c) and (d).

\subsubsection*{Solubility Study}

\begin{figure}[!h]
    \centering
    \includegraphics[width=0.7\linewidth]{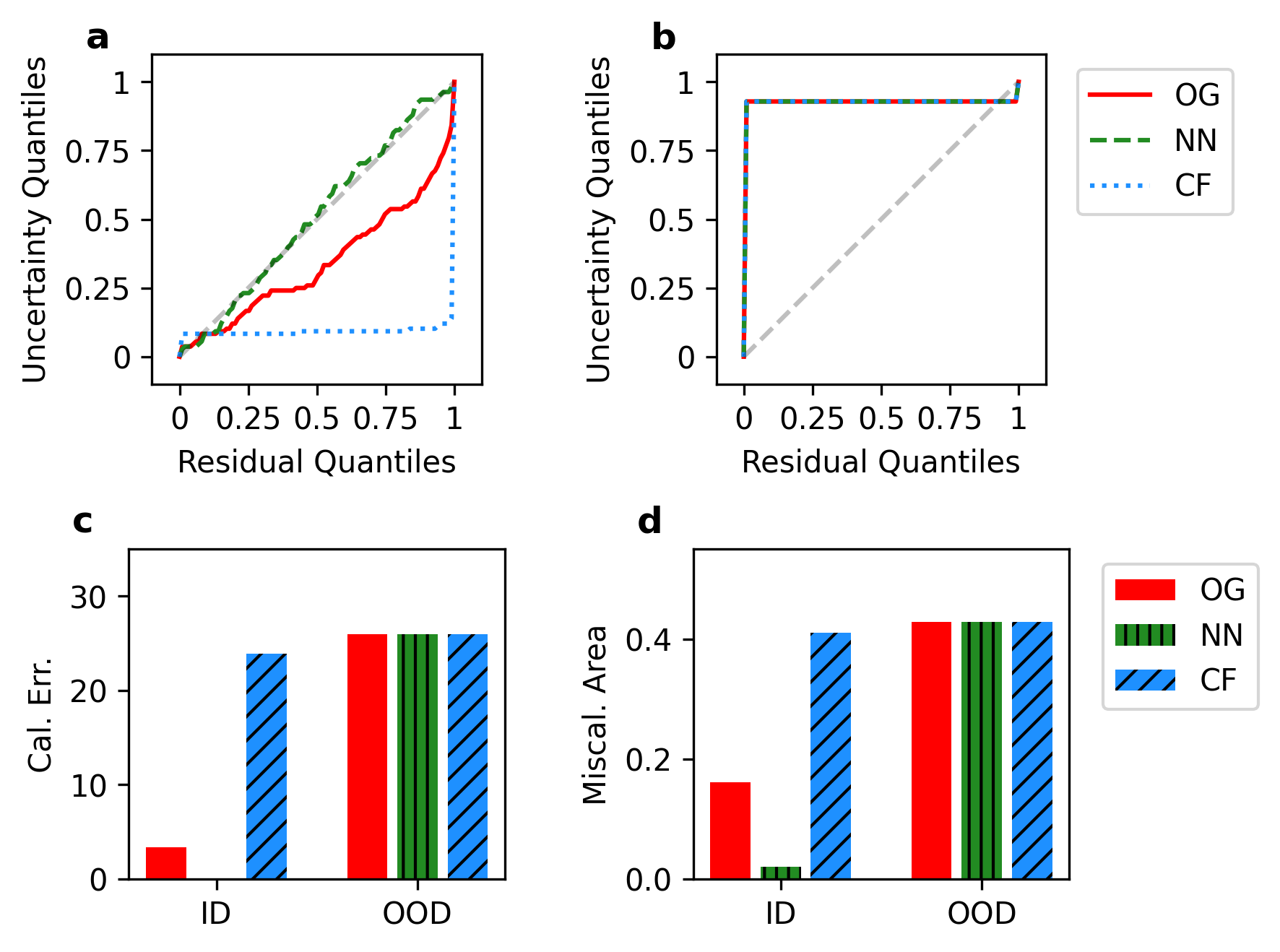}
    \caption{Solubility - Omit outliers - NN}
    \label{fig:esol_uncert_cal}
\end{figure}

\subsubsection*{Bandgap Study}

\begin{figure}[!h]
    \centering
    \includegraphics[width=\linewidth]{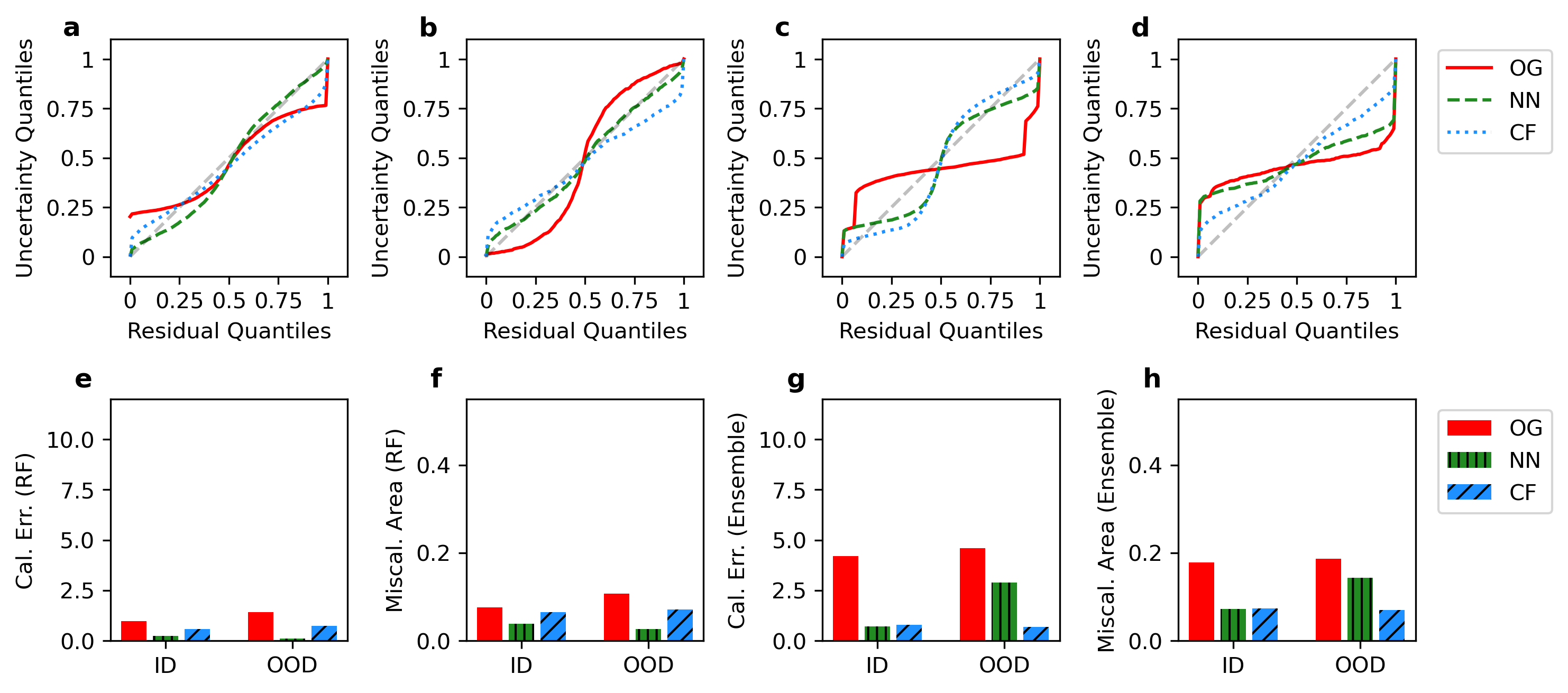}
    \caption{Bandgap - Omit F - QBC:RF}
    \label{fig:calibration_bandgap_F_rf}
\end{figure}

\begin{figure}[!h]
    \centering
    \includegraphics[width=\linewidth]{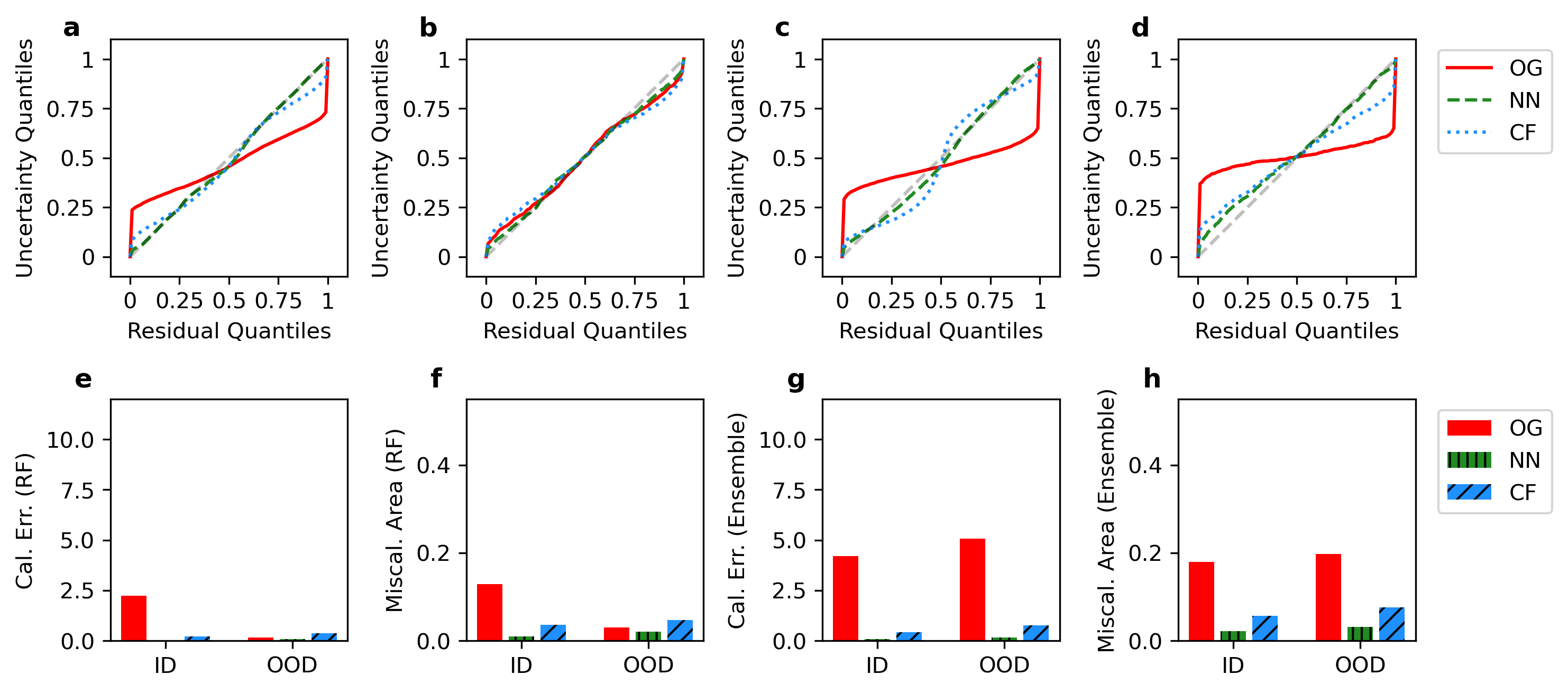}
    \caption{Bandgap - Omit F - QBC:XGB}
    \label{fig:calibration_bandgap_F_xgb}
\end{figure}

\begin{figure}[!h]
    \centering
    \includegraphics[width=\linewidth]{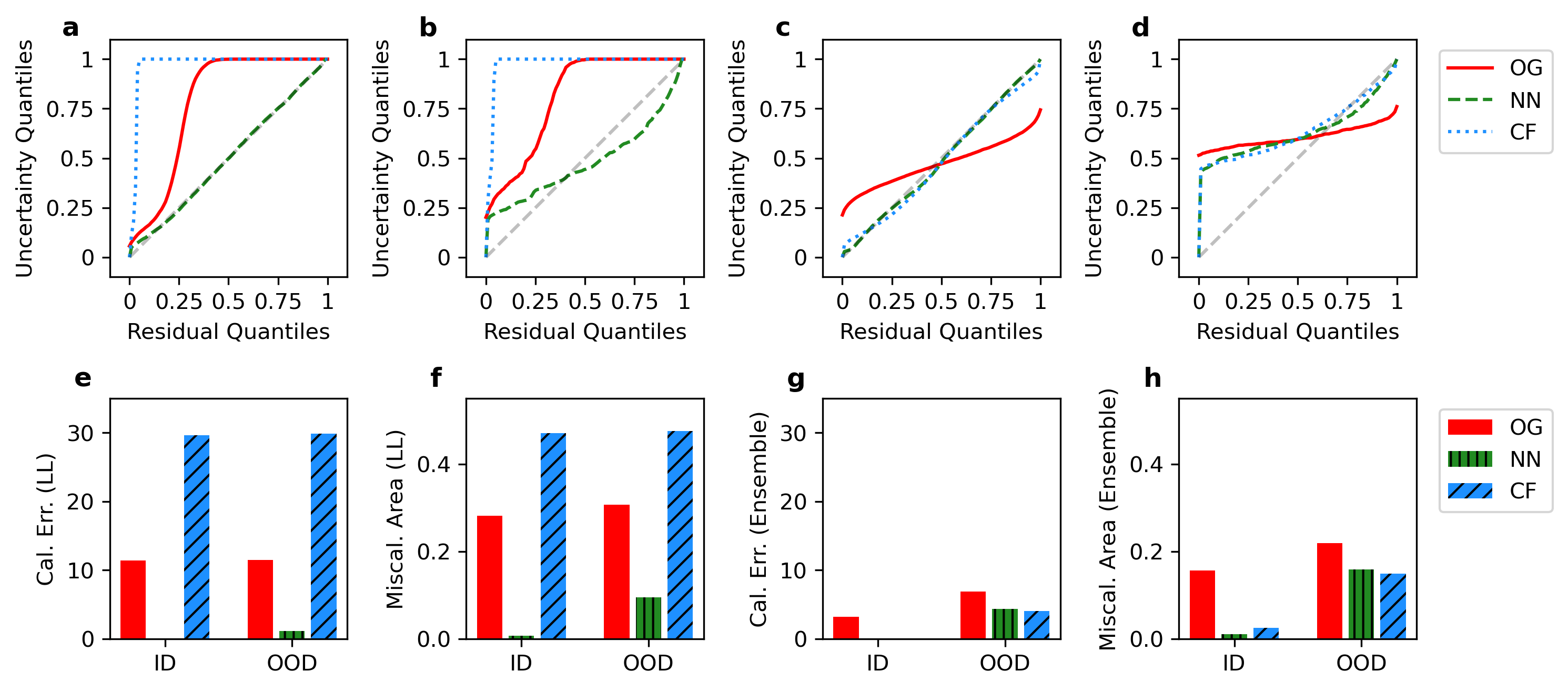}
    \caption{Bandgap - Omit F - ALIGNN Loss Landscape and Random Ensemble}
    \label{fig:calibration_bandgap_F_alignn}
\end{figure}

\begin{figure}[!h]
    \centering
    \includegraphics[width=\linewidth]{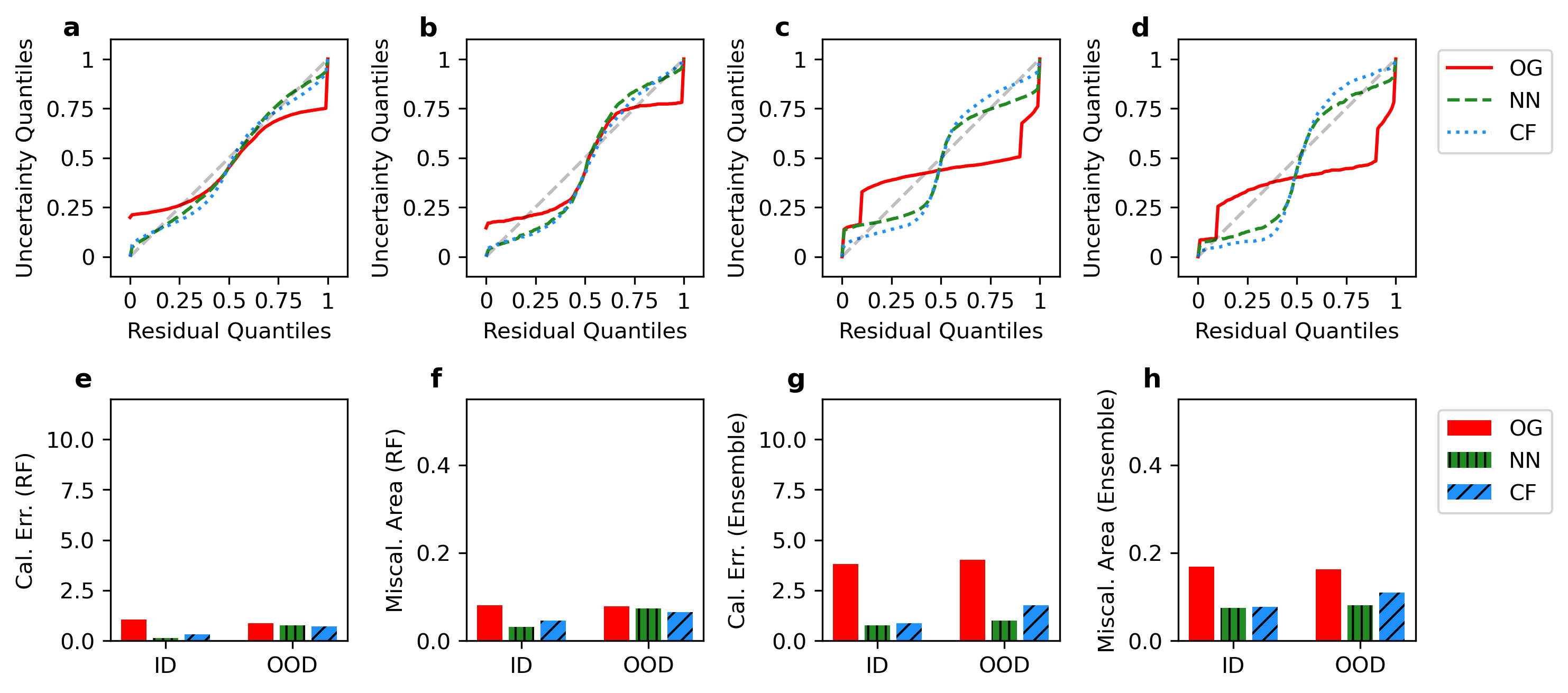}
    \caption{Bandgap – Omit Fe - QBC:RF}
    \label{fig:bgap_Fe_rf_cal_summary}
\end{figure}

\begin{figure}[!h]
    \centering
    \includegraphics[width=\linewidth]{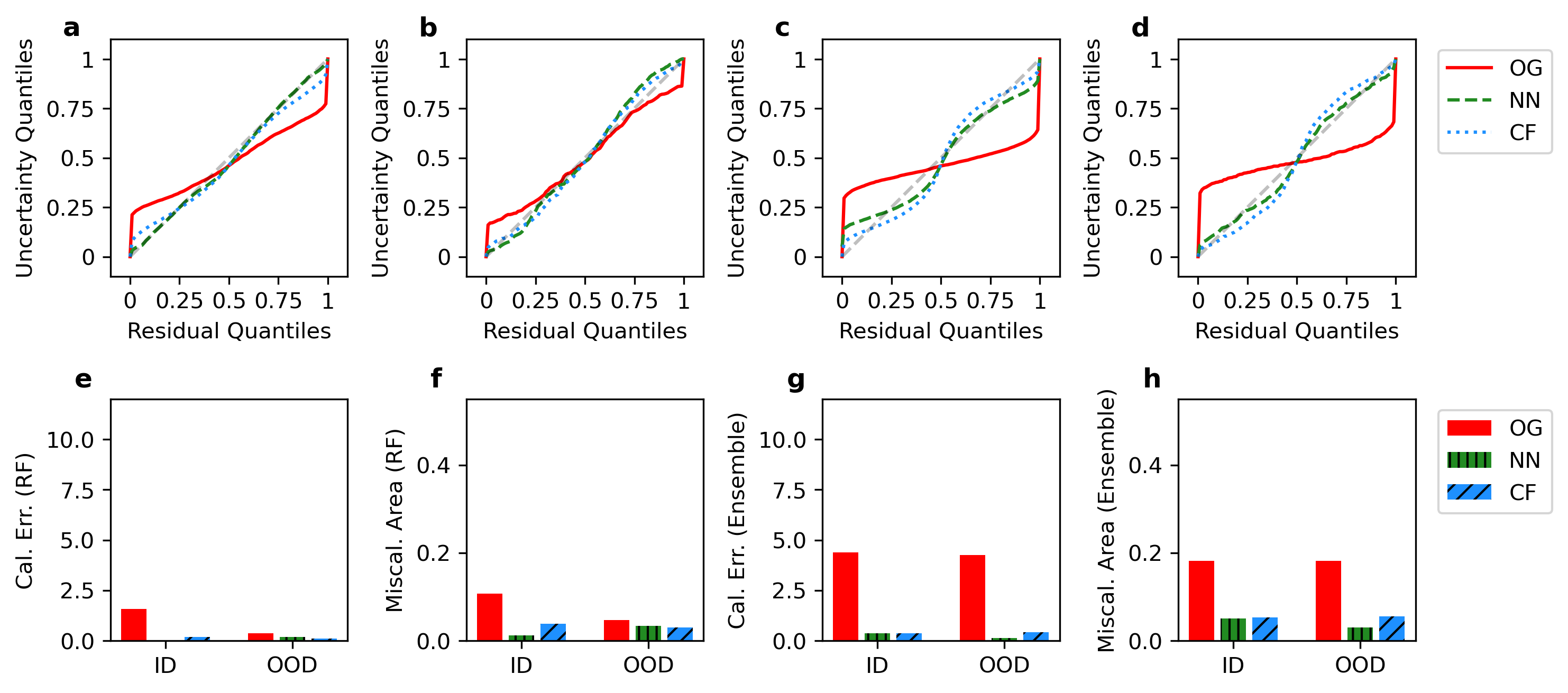}
    \caption{Bandgap - Omit Fe - QBC:XGB}
    \label{fig:calibration_bandgap_Fe_xgb}
\end{figure}

\begin{figure}[!h]
    \centering
    \includegraphics[width=\linewidth]{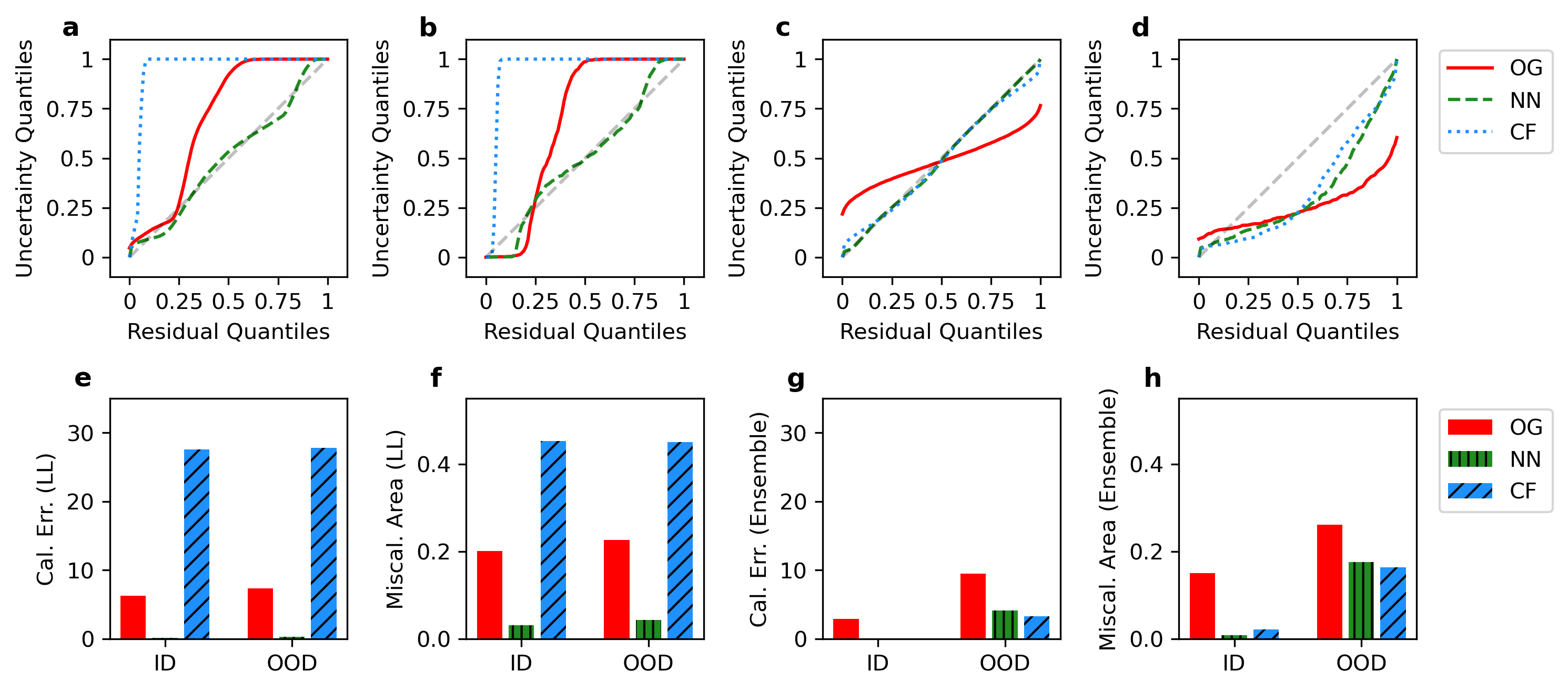}
    \caption{Bandgap - Omit Fe - ALIGNN Loss Landscape and Random Ensemble}
    \label{fig:calibration_bandgap_Fe_alignn}
\end{figure}

\clearpage
\subsubsection*{Formation Energy Study}

\begin{figure}[!h]
    \centering
    \includegraphics[width=\linewidth]{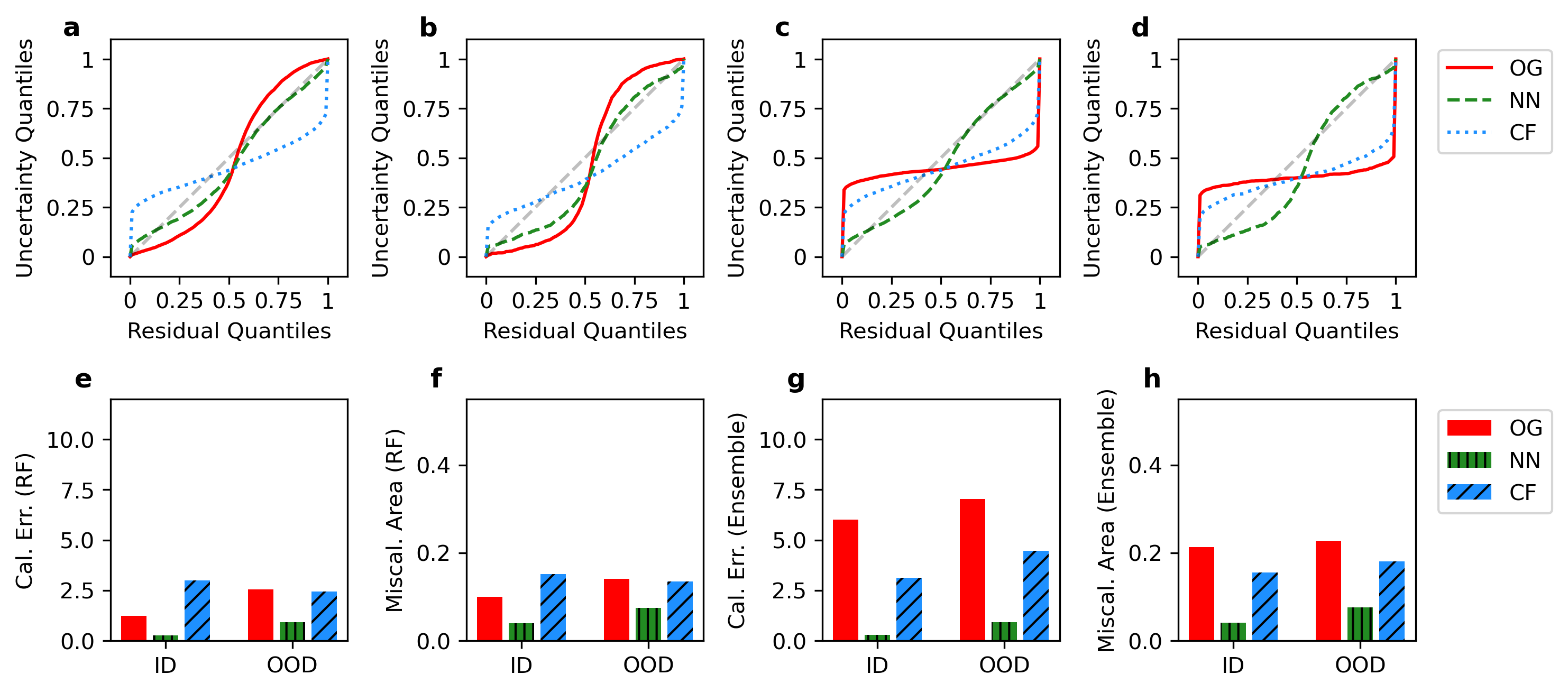}
    \caption{Formation energy - Omit F - QBC:RF. (a) Reliability plot of ID-test 90~\% confidence interval from RF model trained omitting F. (b) Reliability plot of OOD-test 90~\% confidence interval from RF model trained omitting F. (c) Reliability plot of ID-test uncertainty estimates from QBC ensemble. (d) OOD-test uncertainty estimates from QBC ensemble. (e) Calibration errors from (a) and (b). (f) Miscalibration area from (a) and (b). (g) Calibration error from (c) and (d). (h) Miscalibration area from (c) and (d). The uncertainty estimates from the RF model are more calibrated w.r.t. the RF model residual distribution than the QBC ensemble uncertainty estimates.}
    \label{fig:calibration_eform_F_rf}
\end{figure}

\begin{figure}[!h]
    \centering
    \includegraphics[width=\linewidth]{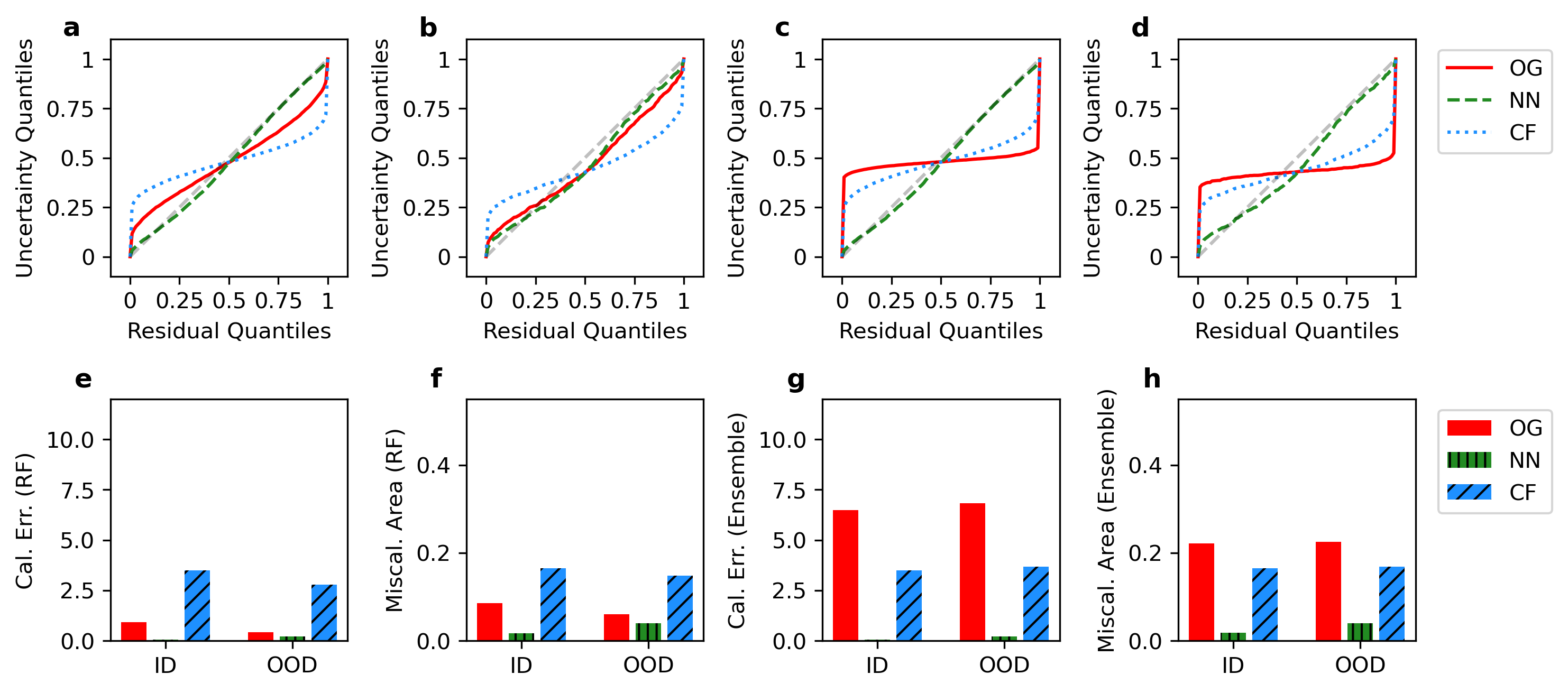}
    \caption{Formation Energy – Omit F – QBC:XGB. (a) ID-test 90~\% confidence interval from XGB model trained omitting F. (b) OOD-test 90~\% confidence interval from XGB model trained omitting F. (c) ID-test uncertainty estimates from QBC ensemble. (d) OOD-test uncertainty estimates from QBC ensemble. (e) Calibration errors from (a) and (b). (f) Miscalibration area from (a) and (b). (g) Calibration error from (c) and (d). (h) Miscalibration area from (c) and (d). The uncertainty estimates from the RF model are more calibrated w.r.t. the RF model residual distribution than the QBC ensemble uncertainty estimates.}
    \label{fig:eform_F_xgb_cal_summary}
\end{figure}

\begin{figure}[!h]
    \centering
    \includegraphics[width=\linewidth]{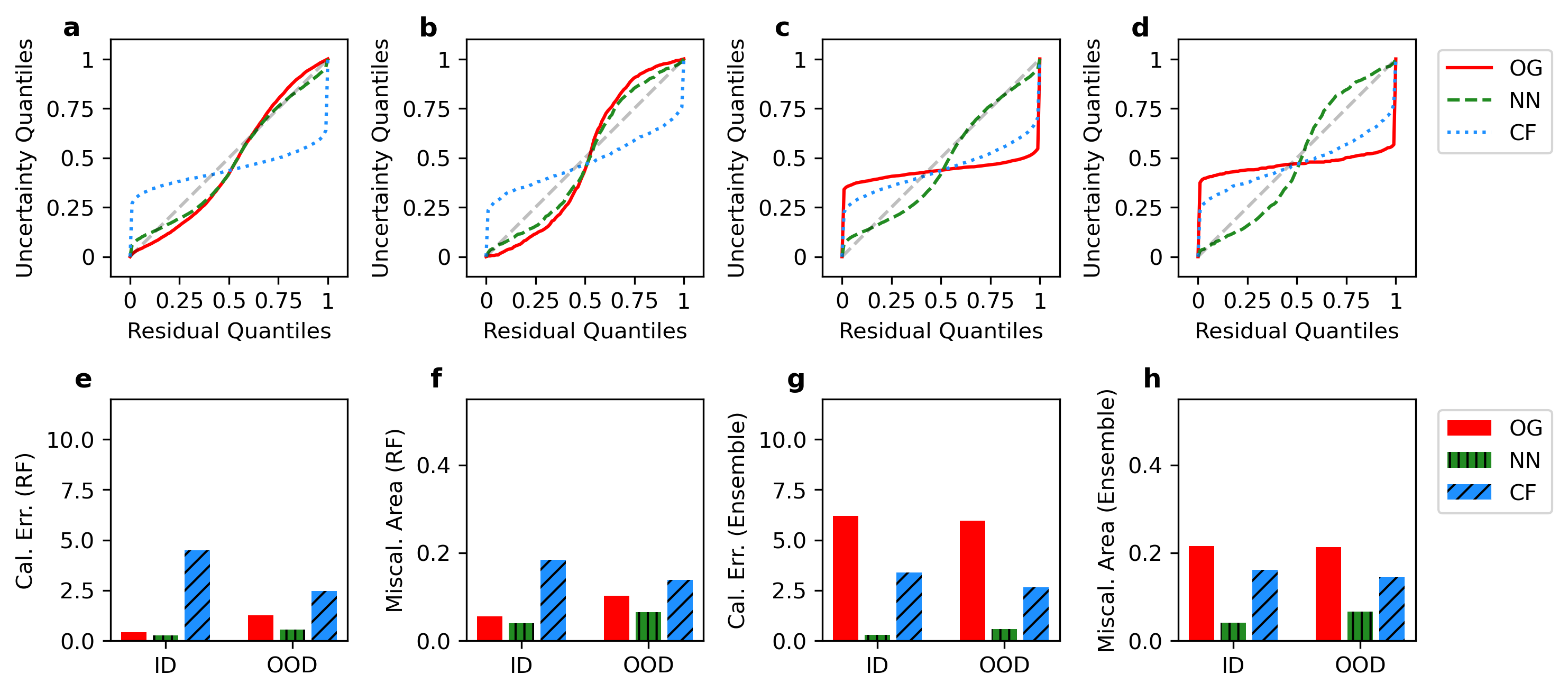}
    \caption{Formation energy - Omit Fe - QBC:RF. (a) ID-test 90~\% confidence interval from RF model trained omitting Fe. (b) OOD-test 90~\% confidence interval from RF model trained omitting Fe. (c) ID-test uncertainty estimates from QBC ensemble. (d) OOD-test uncertainty estimates from QBC ensemble. (e) Calibration errors from (a) and (b). (f) Miscalibration area from (a) and (b). (g) Calibration error from (c) and (d). (h) Miscalibration area from (c) and (d). The uncertainty estimates from the RF model are more calibrated w.r.t. the RF model residual distribution than the QBC ensemble uncertainty estimates.}
    \label{fig:calibration_eform_Fe_rf}
\end{figure}

\begin{figure}[!h]
    \centering
    \includegraphics[width=\linewidth]{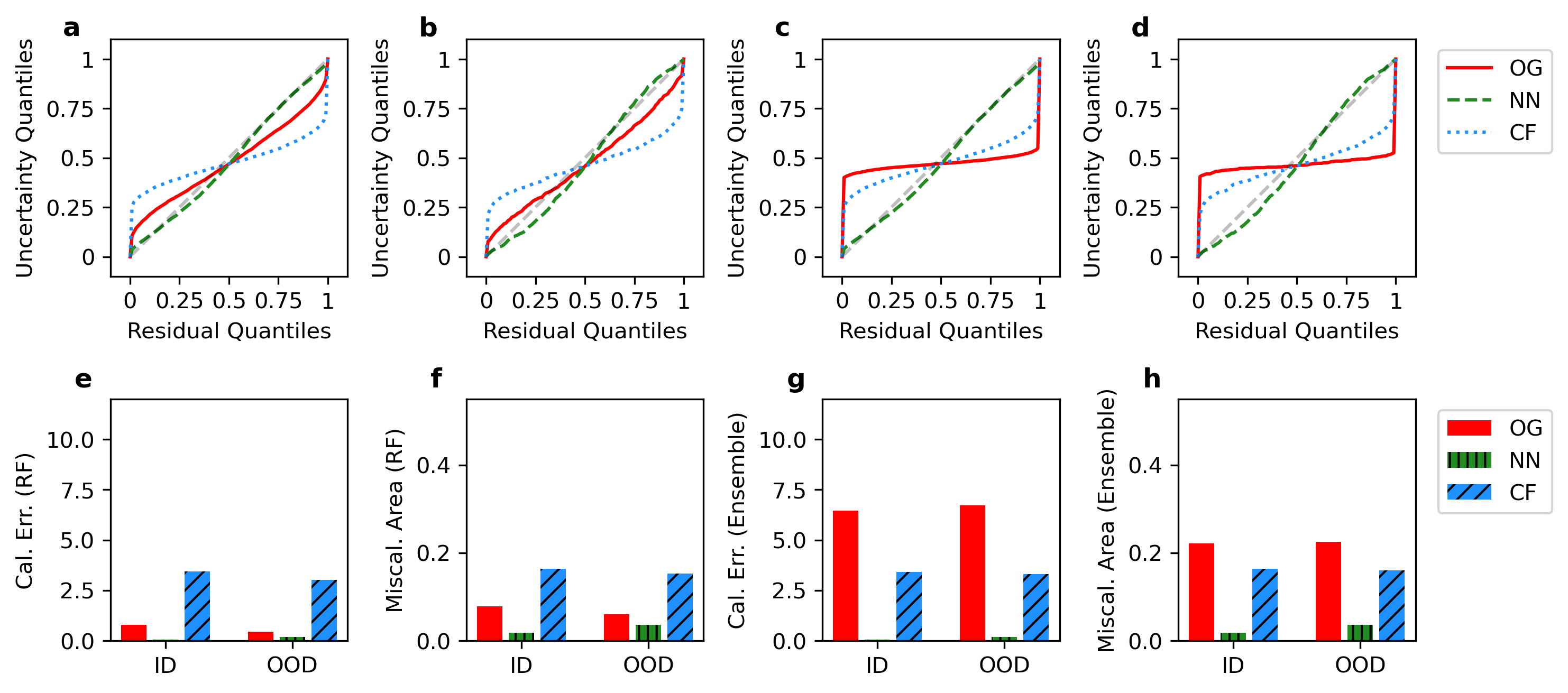}
    \caption{Formation Energy – Omit Fe – QBC:XGB. (a) ID-test 90~\% confidence interval from XGB model trained omitting Fe. (b) OOD-test 90~\% confidence interval from XGB model trained omitting Fe. (c) ID-test uncertainty estimates from QBC ensemble. (d) OOD-test uncertainty estimates from QBC ensemble. (e) Calibration errors from (a) and (b). (f) Miscalibration area from (a) and (b). (g) Calibration error from (c) and (d). (h) Miscalibration area from (c) and (d). The uncertainty estimates from the RF model are more calibrated w.r.t. the RF model residual distribution than the QBC ensemble uncertainty estimates.}
    \label{fig:eform_Fe_xgb_cal_summary}
\end{figure}

\clearpage
\subsection*{Recalibration Models}

When training a recalibration model, the ID-test data and uncertainties are used exclusively; the OOD-test data and uncertainties are reserved for testing.

The first recalibration model used to calibrate uncertainties is a fully connected neural network with three layers [2, 32, 1] and a ReLU activation functions. The loss function minimized during training is an approximation of the calibration error of the reliability plot. Histograms are non-differentiable, leading to an approximation of histogram binning using a high temperature sigmoid function.

\begin{lstlisting}
def custom_calibration_loss(residuals_tensor, pred_stddevs, mu, sigma):

    predicted_pi_tensor = 
        torch.linspace(0, 1, 102, dtype=residuals_tensor.dtype, 
        device=residuals_tensor.device, requires_grad=True)
    
    upper_bounds = normal_ppf(predicted_pi_tensor, mu, sigma) 

    normalized_residuals_tensor = 
        torch.div(residuals_tensor, pred_stddevs+1e-8).squeeze(1)

    # Vectorized approach 
    tmp = normalized_residuals_tensor.unsqueeze(0)
    bounds = upper_bounds[1:-1].unsqueeze(1)

    # Use sigmoid as differentiable approximation to step function
    # High temperature makes it closer to hard threshold
    temperature = 50.0
    soft_counts = 
        torch.sigmoid(temperature * (bounds - tmp))
    densities = 
        soft_counts.mean(dim=1)

    loss = torch.sum((predicted_pi_tensor[1:-1] - densities) ** 2)
    
    return loss
\end{lstlisting}

\begin{figure}[!h]
    \centering
    \includegraphics[width=\linewidth]{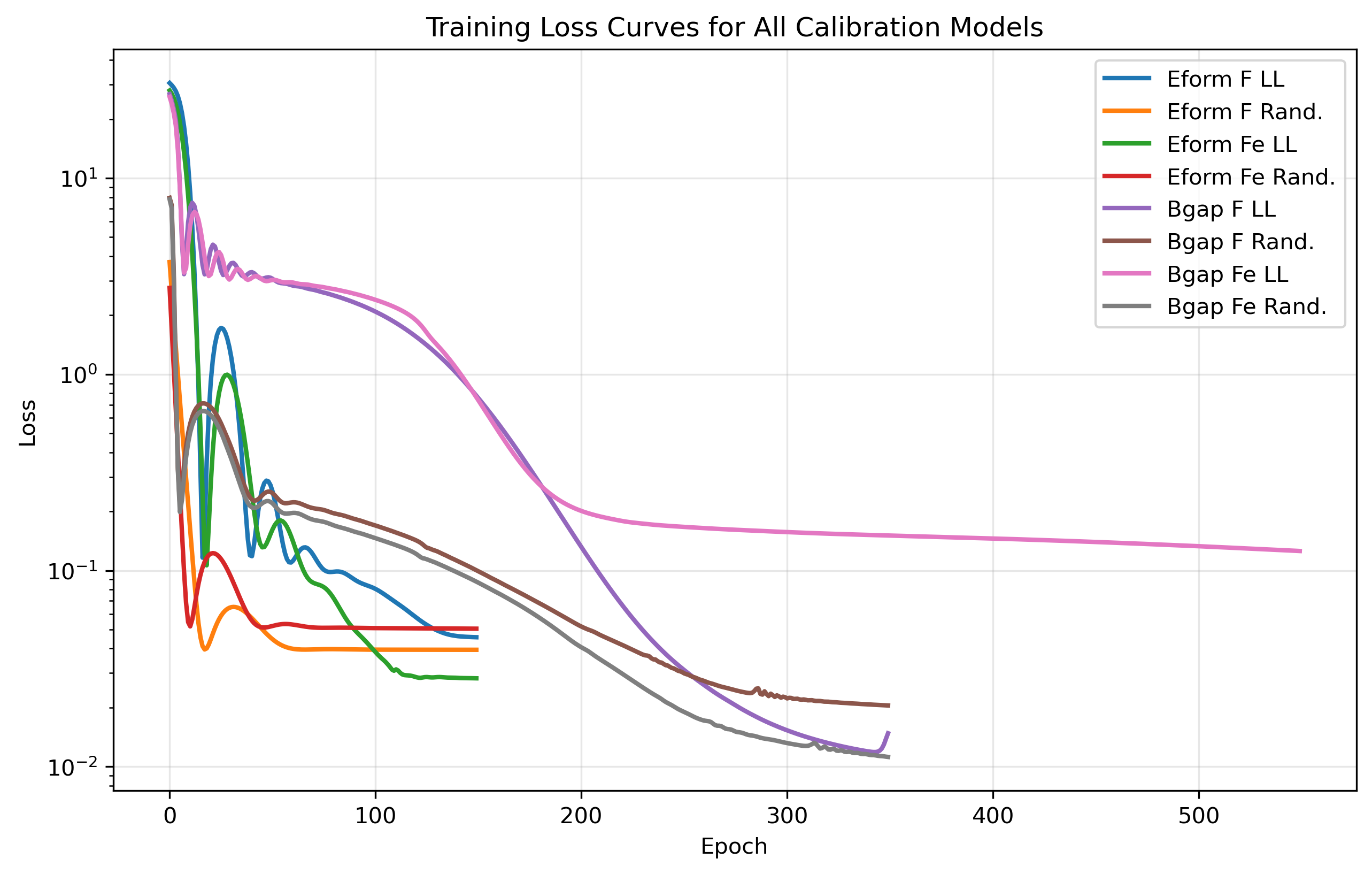}
    \caption{Training losses from recalibration model for LL ensembles and random ensembles.}
    \label{fig:recal_model}
\end{figure}

The calibration model is trained on the entire set of residuals and initial uncertainty estimate (batch size = ($N$ samples, 2) )with a learning rate of $3e-2$. Convergence depends on the random seed and the number of epochs as shown in Figure \ref{fig:recal_model}

Calibration factors were calculated using
% ~\cite{}. 
~\citet{palmer2022calibration}.
This method returns a linear scaling so that
\begin{equation}
    \sigma_{new} = \alpha\cdot\sigma + \beta
\end{equation}
where $\alpha$ and $\beta$ are the calibration factors adjusting the initial uncertainty estimate $\sigma$.  

\begin{figure}
    \centering
    \includegraphics[width=\linewidth]{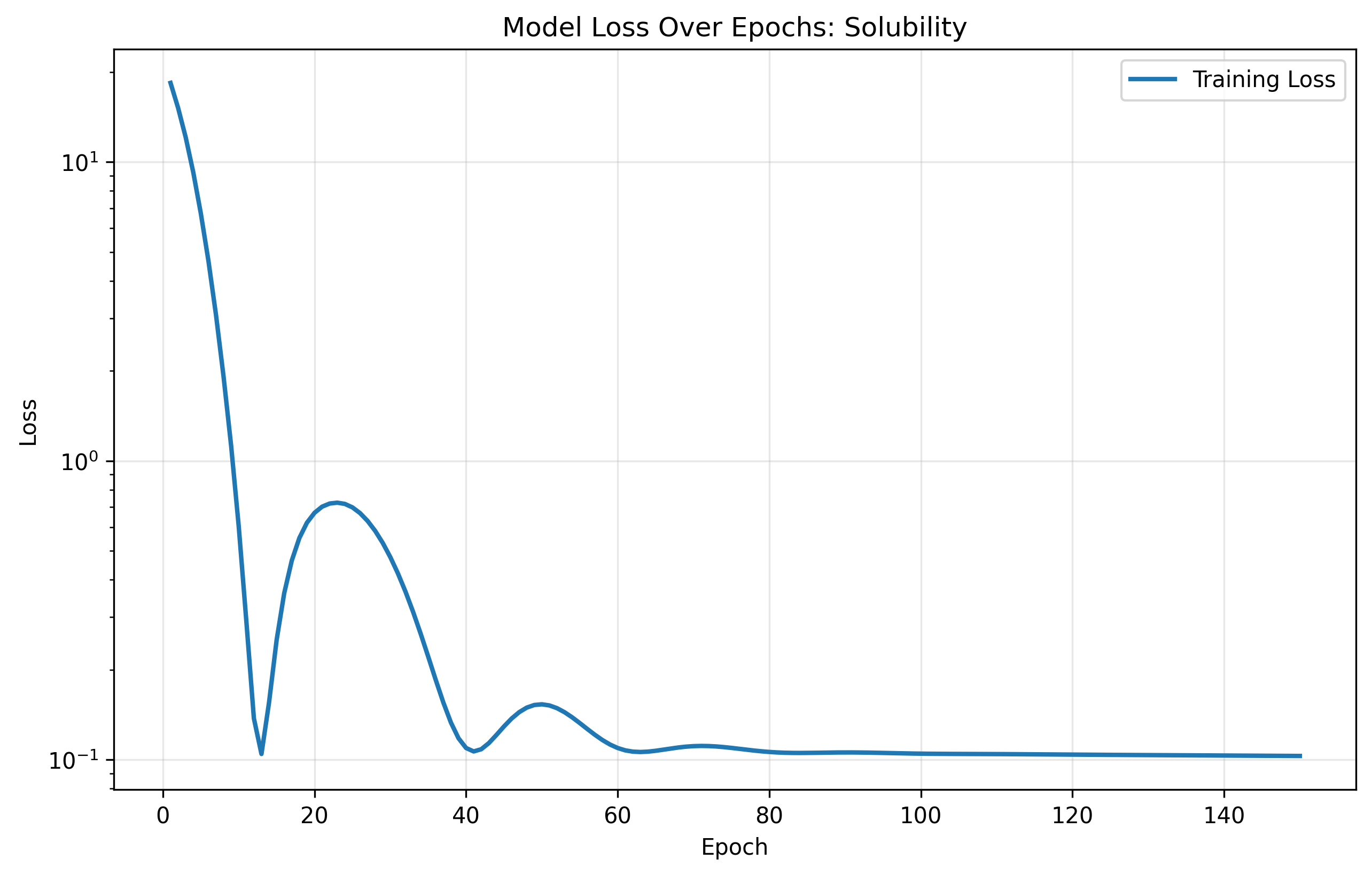}
    \caption{Training loss from recalibration model for LL ensemble and solubility data uncertainty.}
    \label{fig:recal_model_solubility}
\end{figure}

\clearpage
\subsection*{Residuals and Standard Deviation Predictions After Recalibration}

%%%
\subsubsection*{Solubility Study}

\begin{figure}[!h]
    \centering
    \includegraphics[width=\linewidth]{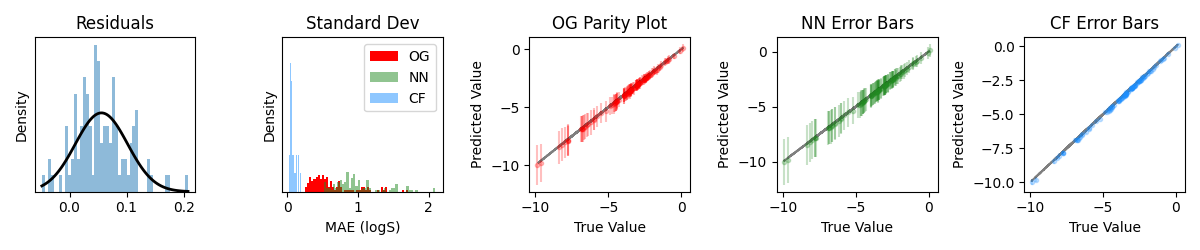}
    \caption{Solubility - Omit Outliers - ID Test - NN Loss Landscape}
    \label{fig:solu_id_alignn}
\end{figure}

\begin{figure}[!h]
    \centering
    \includegraphics[width=\linewidth]{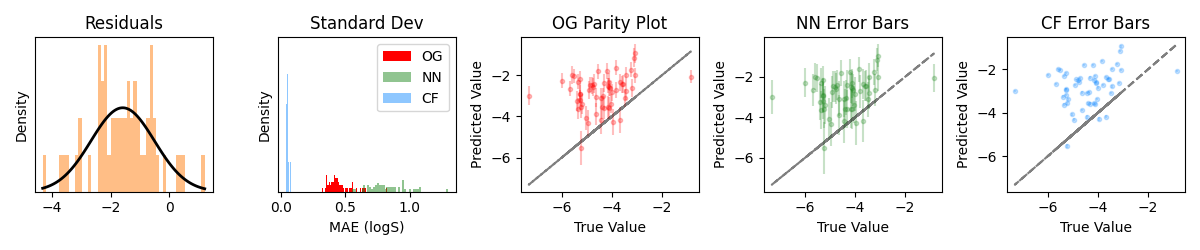}
    \caption{Solubility - Omit Outliers - OOD Test - NN Loss Landscape}
    \label{fig:solu_ood_alignn}
\end{figure}

\subsubsection*{Bandgap Study}

\begin{figure}[!h]
    \centering
    \includegraphics[width=\linewidth]{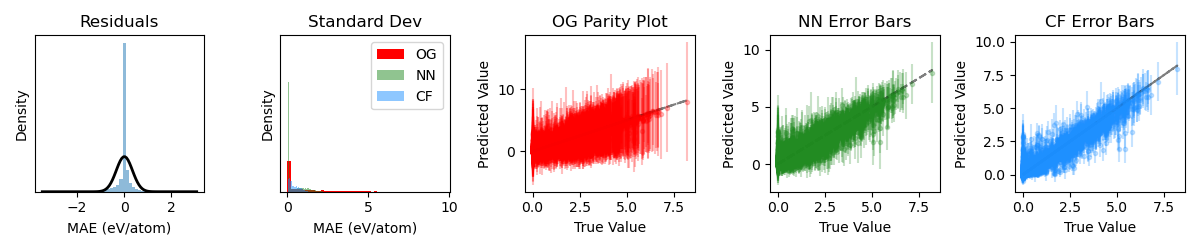}
    \caption{Bandgap - Omit F - ID Test - QBC:RF}
    \label{fig:bgap_F_id_rf_qbc}
\end{figure}

\begin{figure}[!h]
    \centering
    \includegraphics[width=\linewidth]{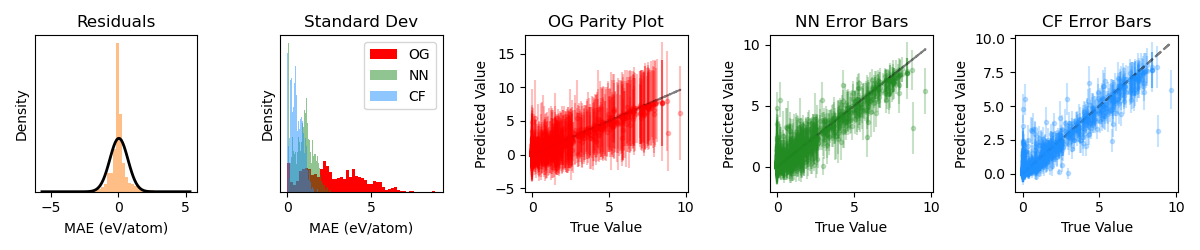}
    \caption{Bandgap - Omit F - OOD Test - QBC:RF}
    \label{fig:bgap_F_ood_rf_qbc}
\end{figure}

\begin{figure}[!h]
    \centering
    \includegraphics[width=\linewidth]{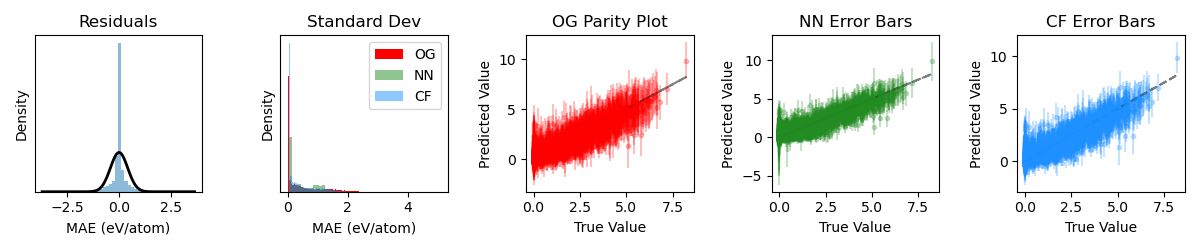}
    \caption{Bandgap - Omit F - ID Test - QBC:XGB}
    \label{fig:bgap_F_id_rf_qbc_xgb}
\end{figure}

\begin{figure}[!h]
    \centering
    \includegraphics[width=\linewidth]{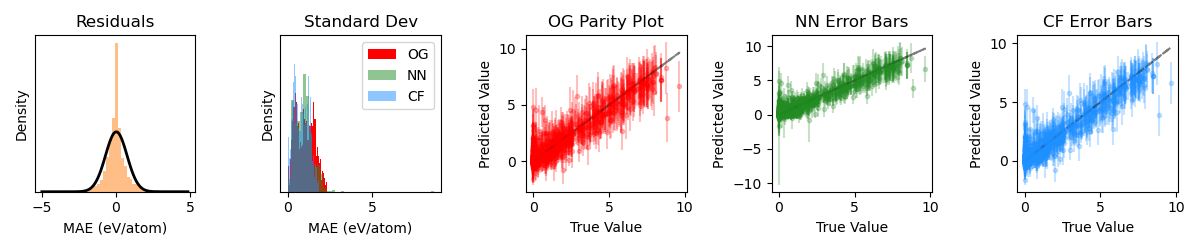}
    \caption{Bandgap - Omit F - OOD Test - QBC:XGB}
    \label{fig:bgap_F_ood_rf_qbc_xgb}
\end{figure}

\begin{figure}[!h]
    \centering
    \includegraphics[width=\linewidth]{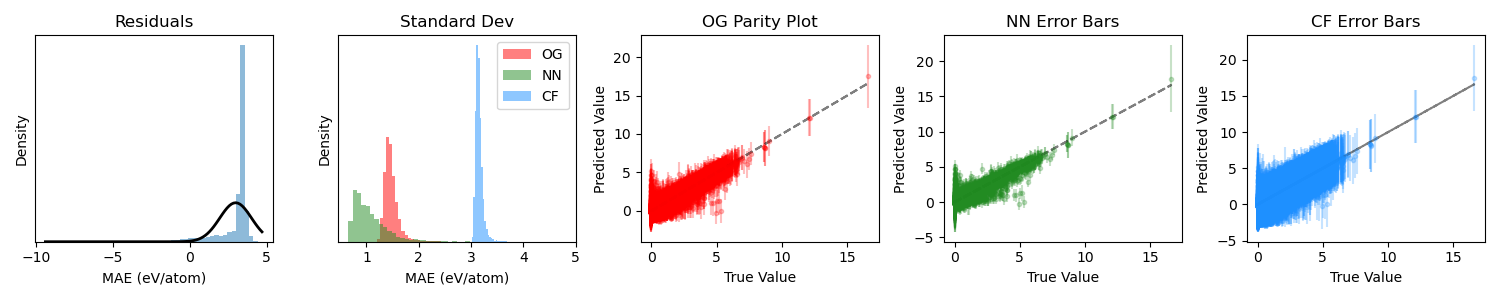}
    \caption{Bandgap - Omit F - ID Test - ALIGNN Loss Landscape}
    \label{fig:bgap_F_id_recal_ll}
\end{figure}

\begin{figure}[!h]
    \centering
    \includegraphics[width=\linewidth]{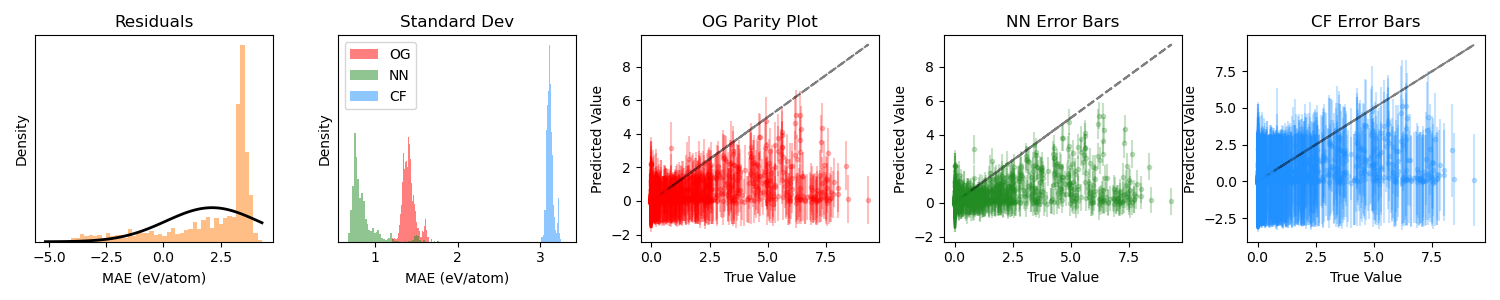}
    \caption{Bandgap - Omit F - OOD Test - ALIGNN Loss Landscape}
    \label{fig:bgap_F_ood_recal_ll}
\end{figure}

\begin{figure}[!h]
    \centering
    \includegraphics[width=\linewidth]{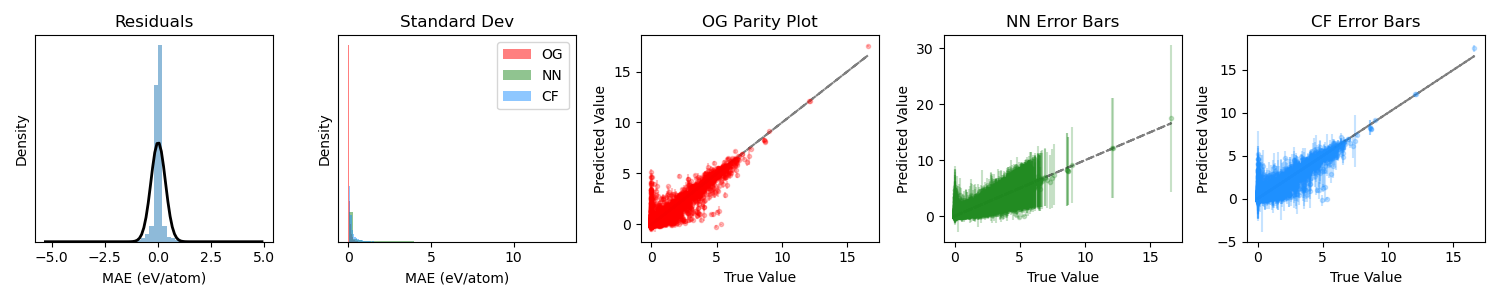}
    \caption{Bandgap - Omit F - ID Test - ALIGNN Random Ensemble}
    \label{fig:bgap_F_id_recal_ensem}
\end{figure}

\begin{figure}[!h]
    \centering
    \includegraphics[width=\linewidth]{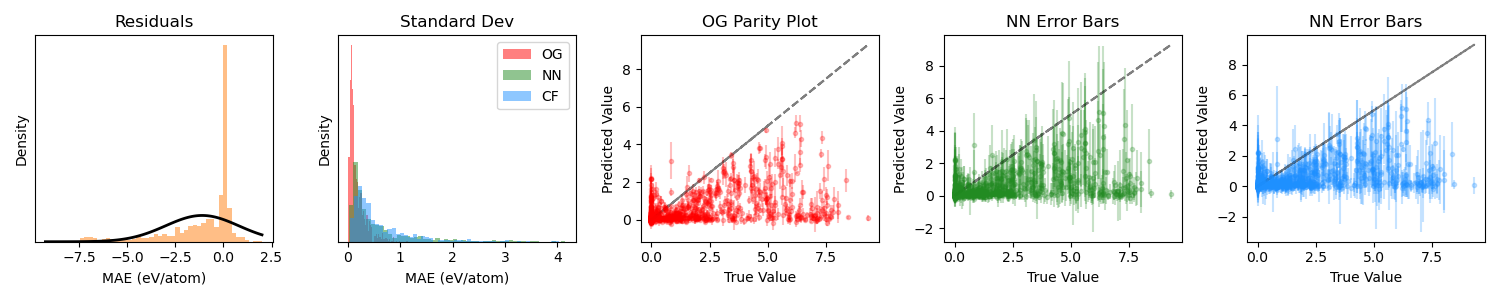}
    \caption{Bandgap - Omit F - OOD Test - ALIGNN Random Ensemble}
    \label{fig:bgap_F_ood_recal_ensem}
\end{figure}

\begin{figure}[!h]
    \centering
    \includegraphics[width=\linewidth]{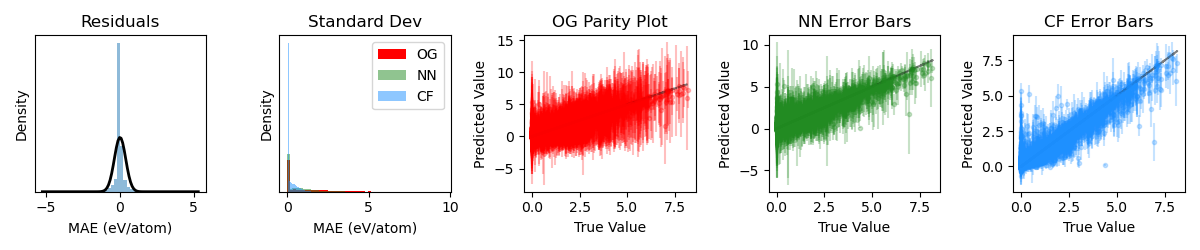}
    \caption{Bandgap - Omit Fe - ID Test - QBC:RF}
    \label{fig:bgap_Fe_id_rf_qbc}
\end{figure}

\begin{figure}[!h]
    \centering
    \includegraphics[width=\linewidth]{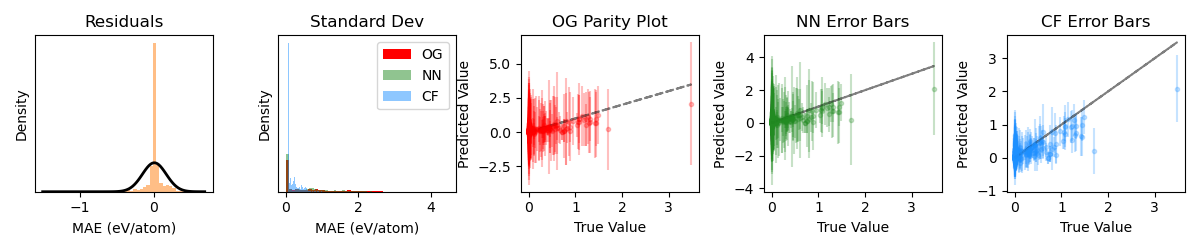}
    \caption{Bandgap - Omit Fe - OOD Test - QBC:RF}
    \label{fig:bgap_Fe_ood_rf_qbc}
\end{figure}

\begin{figure}[!h]
    \centering
    \includegraphics[width=\linewidth]{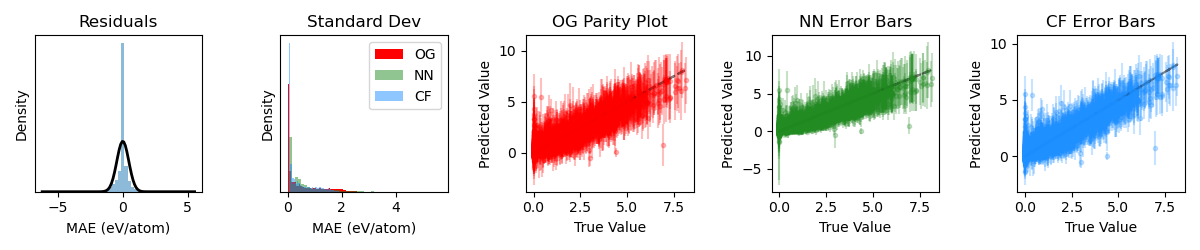}
    \caption{Bandgap - Omit Fe - ID Test - QBC:XGB}
    \label{fig:bgap_Fe_id_xgb_qbc}
\end{figure}

\begin{figure}[!h]
    \centering
    \includegraphics[width=\linewidth]{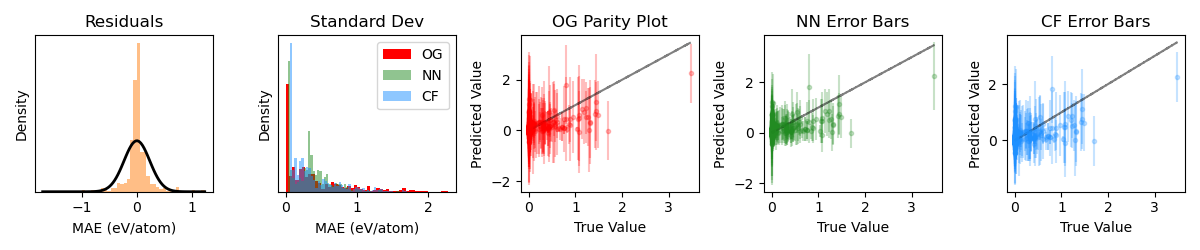}
    \caption{Bandgap - Omit Fe - OOD Test - QBC:XGB}
    \label{fig:bgap_Fe_ood_xgb_qbc}
\end{figure}

\begin{figure}[!h]
    \centering
    \includegraphics[width=\linewidth]{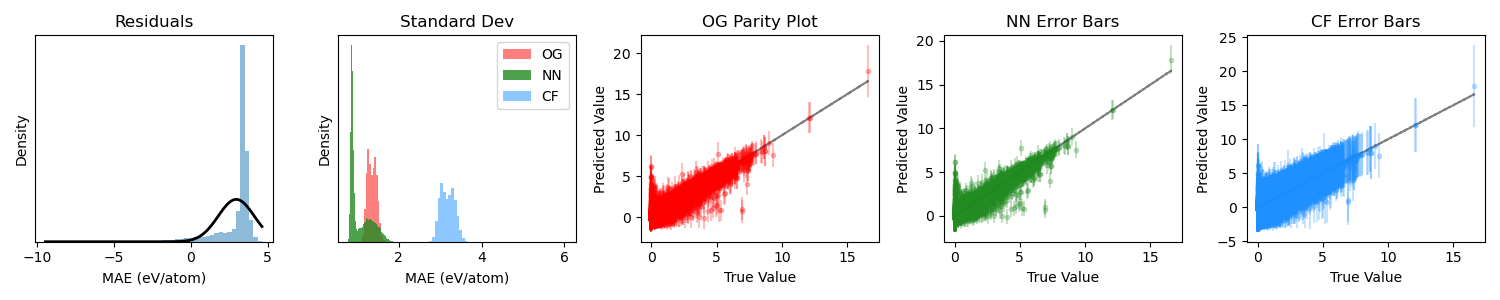}
    \caption{Bandgap - Omit Fe - ID Test - ALIGNN Loss Landscape}
    \label{fig:bgap_Fe_id_recal_ll}
\end{figure}

\begin{figure}[!h]
    \centering
    \includegraphics[width=\linewidth]{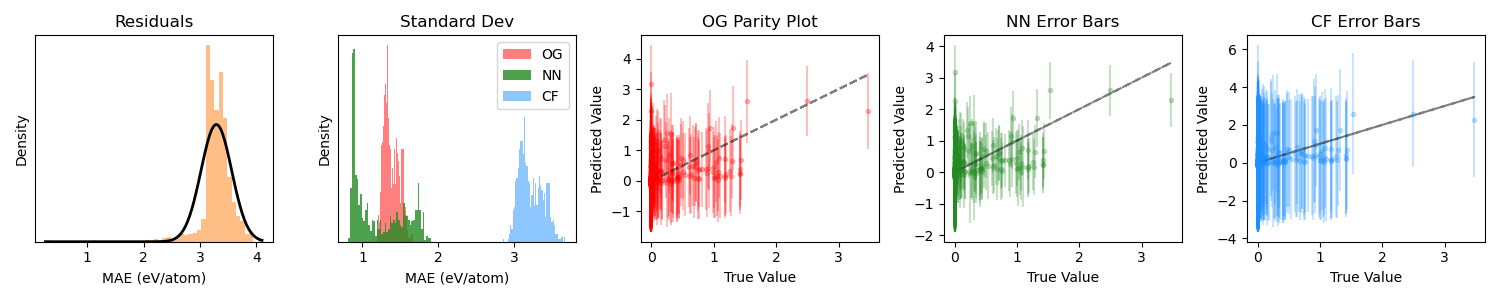}
    \caption{Bandgap - Omit Fe - OOD Test - ALIGNN Loss Landscape}
    \label{fig:bgap_Fe_ood_recal_ll}
\end{figure}

\begin{figure}[!h]
    \centering
    \includegraphics[width=\linewidth]{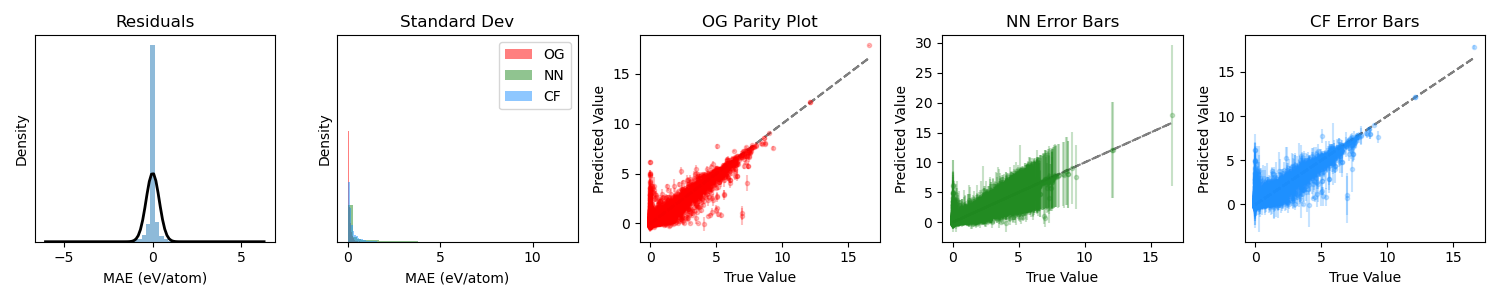}
    \caption{Bandgap - Omit Fe - ID Test - ALIGNN Random Ensemble}
    \label{fig:bgap_Fe_id_recal_ensem}
\end{figure}

\begin{figure}[!h]
    \centering
    \includegraphics[width=\linewidth]{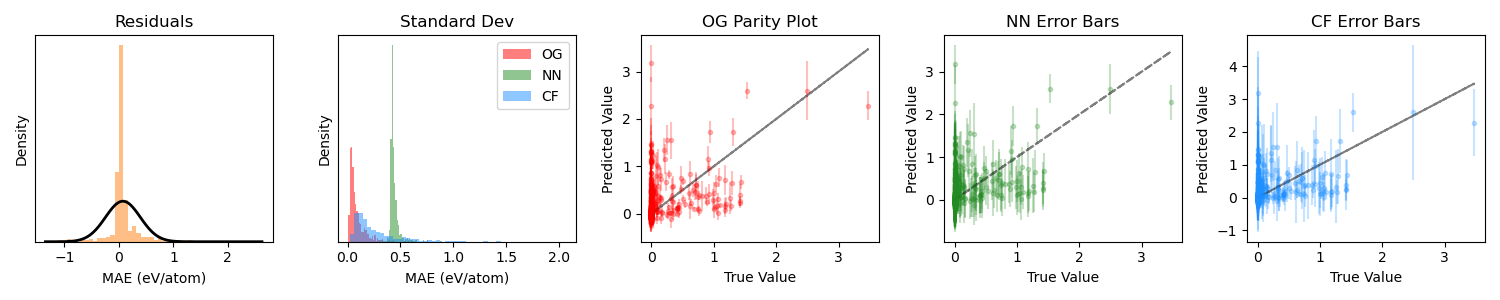}
    \caption{Bandgap - Omit Fe - OOD Test - ALIGNN Random Ensemble}
    \label{fig:bgap_Fe_ood_recal_ensem}
\end{figure}

\clearpage
\subsubsection*{Formation Energy Study}

\begin{figure}[!h]
    \centering
    \includegraphics[width=\linewidth]{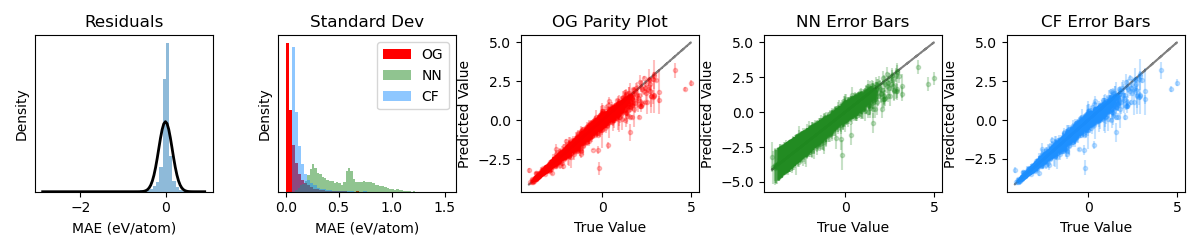}
    \caption{Formation Energy - Omit F - ID Test - QBC:RF}
    \label{fig:eform_F_id_rf_qbc}
\end{figure}

\begin{figure}[!h]
    \centering
    \includegraphics[width=\linewidth]{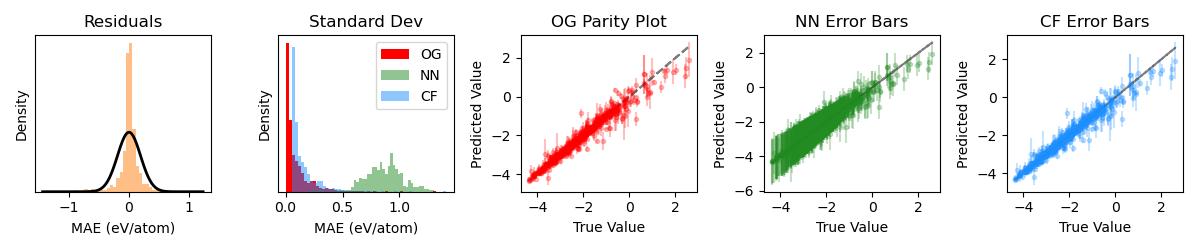}
    \caption{Formation Energy - Omit F - OOD Test - QBC:RF}
    \label{fig:eform_F_ood_rf_qbc}
\end{figure}

\begin{figure}[!h]
    \centering
    \includegraphics[width=\linewidth]{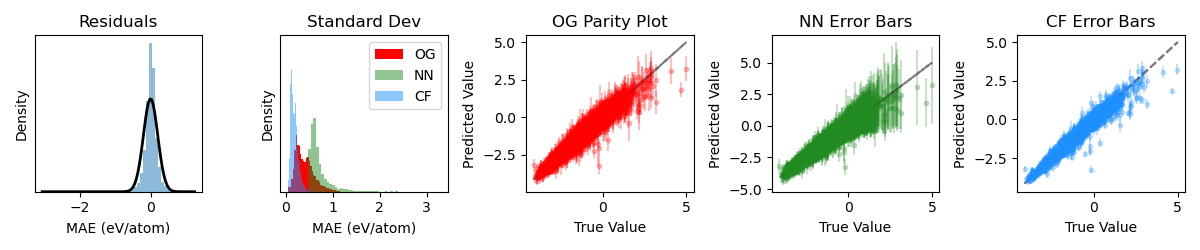}
    \caption{Formation Energy - Omit F - ID Test - QBC:XGB}
    \label{fig:eform_F_id_xgb_qbc}
\end{figure}

\begin{figure}[!h]
    \centering
    \includegraphics[width=\linewidth]{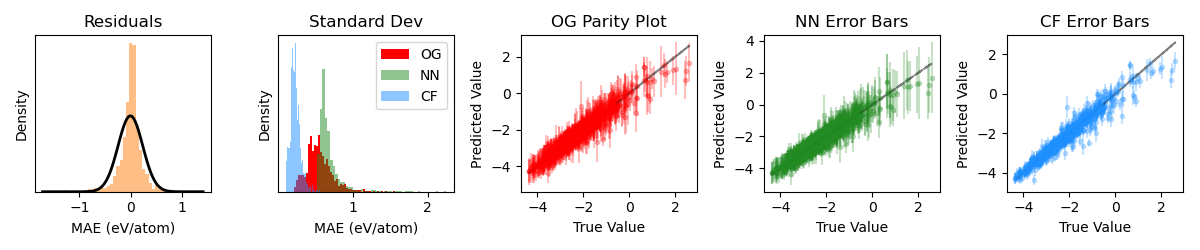}
    \caption{Formation Energy - Omit F - OOD Test - QBC:XGB}
    \label{fig:eform_F_ood_xgb_qbc}
\end{figure}

\begin{figure}[!h]
    \centering
    \includegraphics[width=\linewidth]{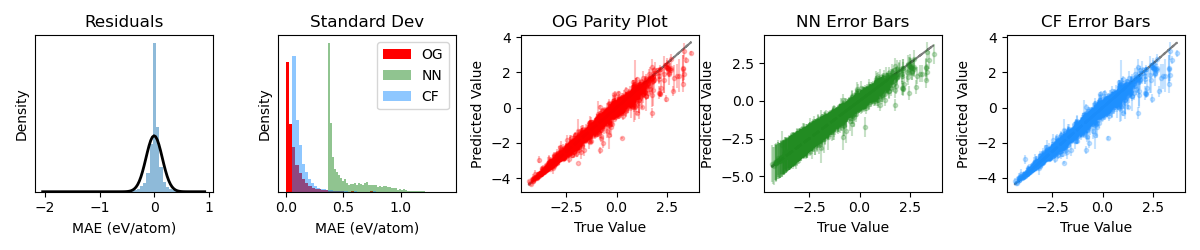}
    \caption{Formation Energy - Omit Fe - ID Test - QBC:RF}
    \label{fig:eform_Fe_id_rf_qbc}
\end{figure}

\begin{figure}[!h]
    \centering
    \includegraphics[width=\linewidth]{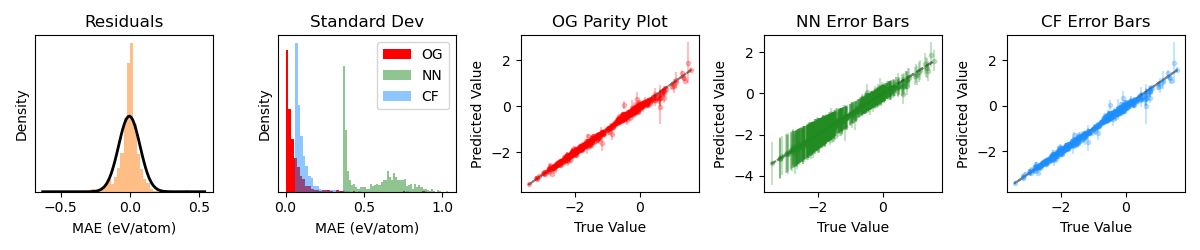}
    \caption{Formation Energy - Omit Fe - OOD Test - QBC:RF}
    \label{fig:eform_Fe_ood_rf_qbc}
\end{figure}

\begin{figure}[!h]
    \centering
    \includegraphics[width=\linewidth]{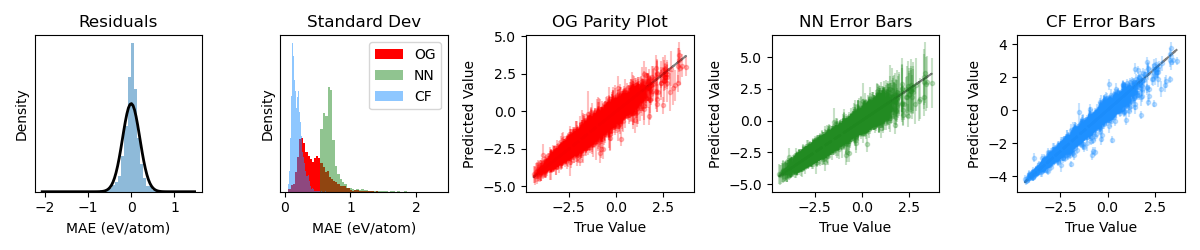}
    \caption{Formation Energy - Omit Fe - ID Test - QBC:XGB}
    \label{fig:eform_Fe_id_xgb_qbc}
\end{figure}

\begin{figure}[!h]
    \centering
    \includegraphics[width=\linewidth]{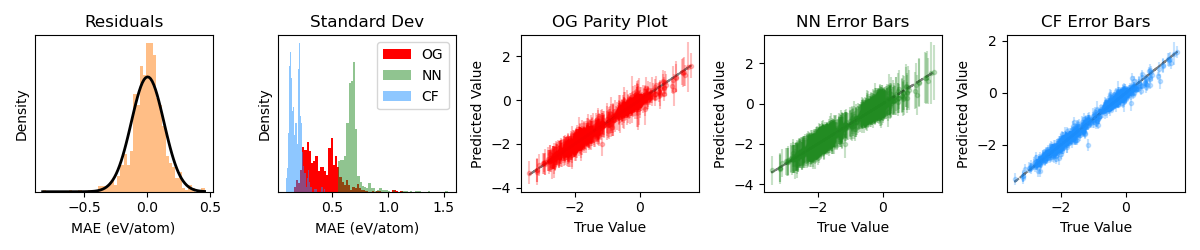}
    \caption{Formation Energy - Omit Fe - OOD Test - QBC:XGB}
    \label{fig:eform_Fe_ood_xgb_qbc}
\end{figure}

\clearpage
\section*{Active Learning Studies}

\begin{figure}[!h]
    \centering
    \includegraphics[width=\linewidth]{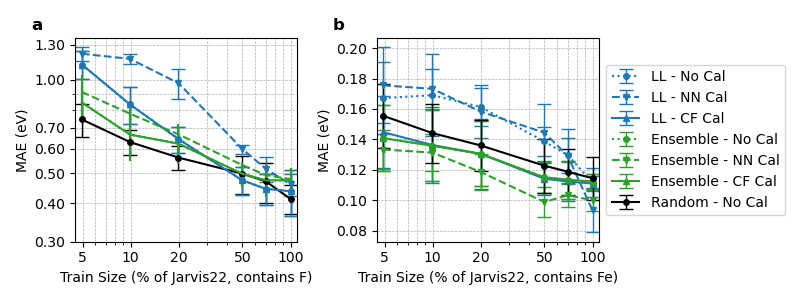}
    \caption{Active learning study for bandgap where (a) the OOD-train set is fluorine-containing compounds and (b) the OOD-train set is iron-containing compounds.}
    \label{fig:active_bgap}
\end{figure}

\begin{figure}[!h]
    \centering
    \includegraphics[width=\linewidth]{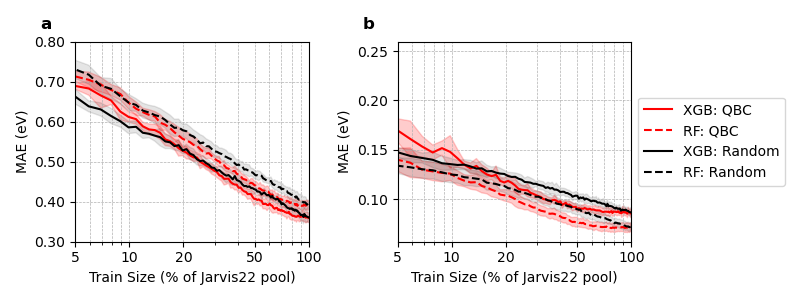}
    \caption{Active learning study for bandgap with query-by-comittee (QBC) for random forest (RF) and XGBoost (XGB) models. (a) Fluorine containing compounds are the OOD-training pool. (b) Iron containing compounds are the OOD-training pool.}
    \label{fig:active_xgb_rf_bgap}
\end{figure}

\clearpage
\section*{Sampling Target Distributions for Active Learning}

\begin{figure}[!h]
    \centering
    \begin{subfigure}[b]{\textwidth}
    \centering
        \includegraphics[width=0.8\textwidth]{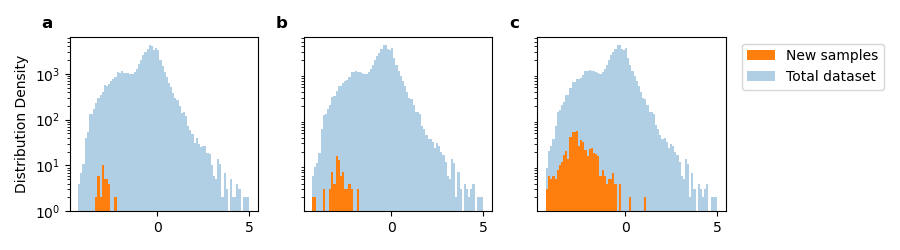}
    \end{subfigure}
    \begin{subfigure}[b]{\textwidth}
    \centering
        \includegraphics[width=0.8\textwidth]{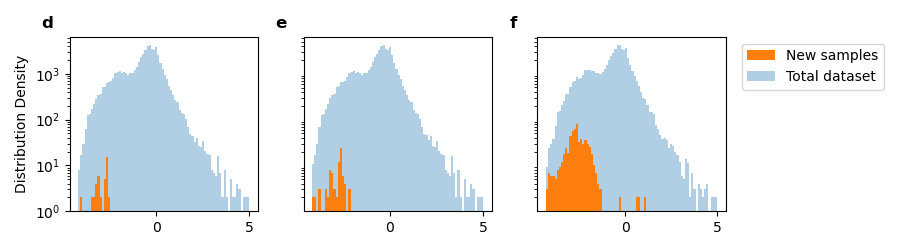}
    \end{subfigure}
    \begin{subfigure}[b]{\textwidth}
    \centering
        \includegraphics[width=0.8\textwidth]{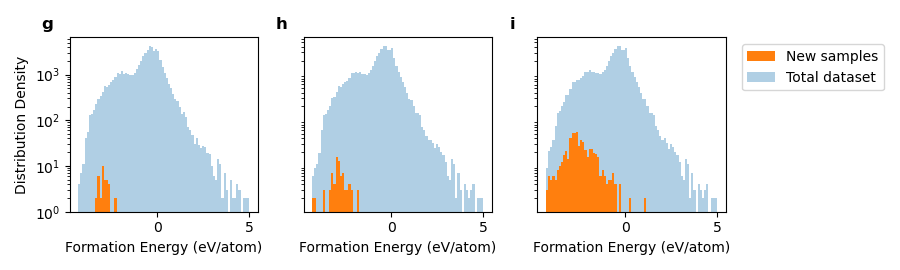}
    \end{subfigure}
    \caption{Differences in the target distribution with uncertainty estimates from the random ensemble using no calibration (a, b, c), neural network calibration (d, e, f), and calibration factors (g, h, i) for the formation energy prediction task when fluorine-containing compounds form the hold-out set. Column 1 introduces 5~\% (43 samples) of the held-out data, column 2 introduces 10~\% (86 samples) of the held-out data, and column 3 introduces 70~\% (607 samples) of the held-out data. The differences in acquisition policies is subtle.}
    \label{fig:sample_uncertainty_dist}
\end{figure}

\begin{figure}
    \centering
    \includegraphics[width=\linewidth]{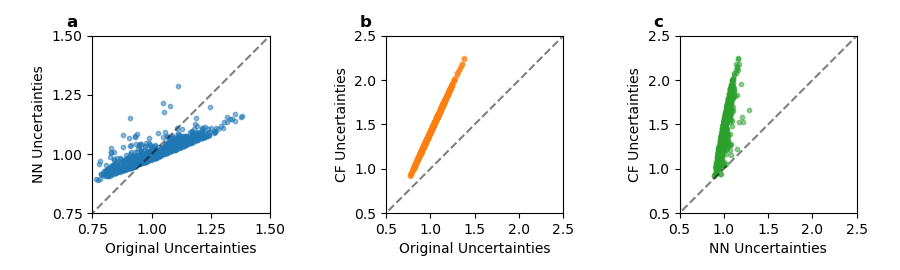}
    \caption{Comparison of how calibration methods transform uncertainties from the loss landscape ensemble trained for formation energy omit F task. (a) The values calibrated with the neural network, (b) values calibrated with calibration factors, and (c) a comparison of the calibration factors to the neural network calibration. The dashed line indicates where the two transforms are equivalent.}
    \label{fig:sample_ll_eform_F_comp_uncert}
\end{figure}

\begin{figure}[!h]
    \centering
    \begin{subfigure}[b]{\textwidth}
        \includegraphics[width=\textwidth]{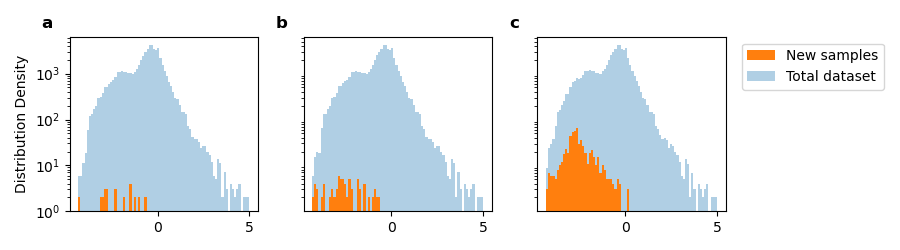}
    \end{subfigure}
    \begin{subfigure}[b]{\textwidth}
        \includegraphics[width=\textwidth]{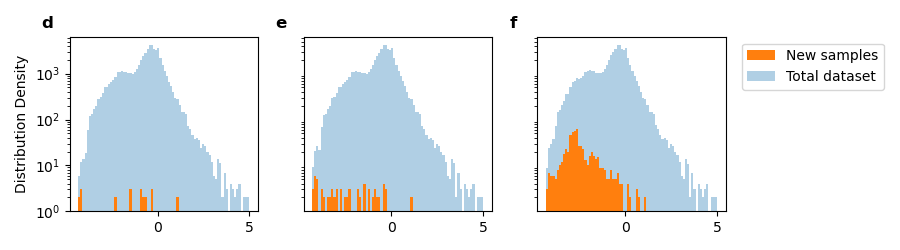}
    \end{subfigure}
    \begin{subfigure}[b]{\textwidth}
        \includegraphics[width=\textwidth]{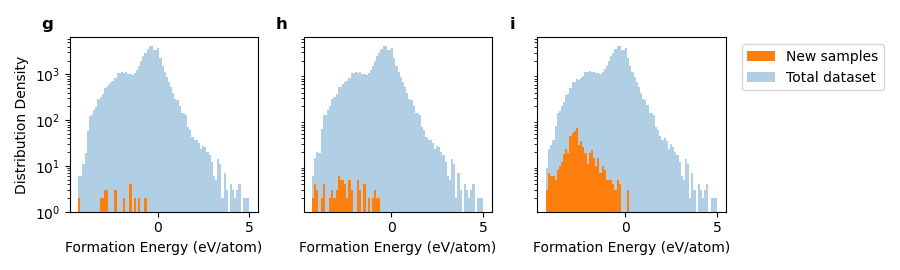}
    \end{subfigure}
    \caption{Using loss landscape uncertainties for active learning the formation energy omit F task. (a, b, c) Random sampling from uncalibrated loss landscapes. (d, e, f) Uncertainty calibrated using neural network. Column 1: 5~\% of the held out training data. Column 2: 10~\% of the held out training data. Column 3: 70~\% of the held out training data.}
    \label{fig:active_learning_dist_eform_F_ll}
\end{figure}

\begin{figure}
    \centering
    \includegraphics[width=\linewidth]{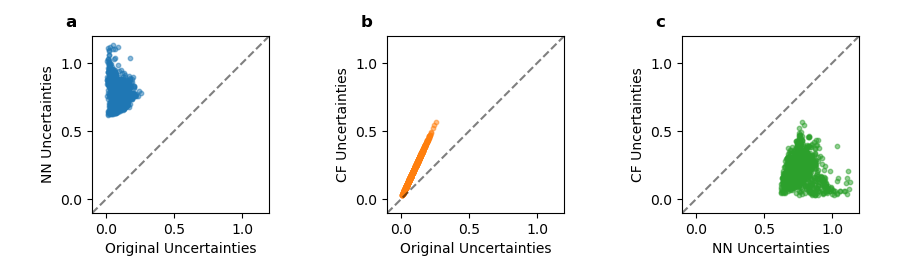}
    \caption{Comparison of how calibration methods transform uncertainties from the random ensemble trained for formation energy omit F task. (a) The values calibrated with the neural network, (b) values calibrated with calibration factors, and (c) a comparison of the calibration factors to the neural network calibration. The dashed line indicates where the two transforms are equivalent.}
    \label{fig:sample_ensem_eform_F_comp_uncert}
\end{figure}

\begin{figure}[!h]
    \centering
    \begin{subfigure}[b]{\textwidth}
        \includegraphics[width=\textwidth]{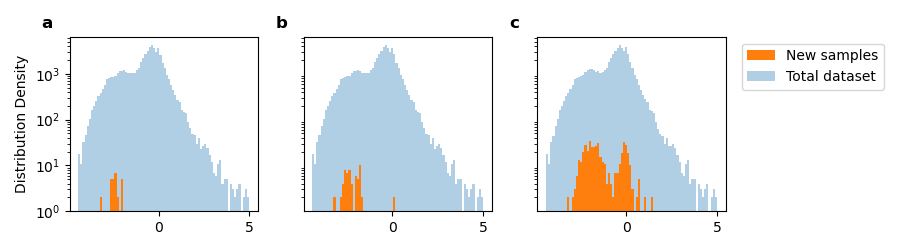}
    \end{subfigure}
    \begin{subfigure}[b]{\textwidth}
        \includegraphics[width=\textwidth]{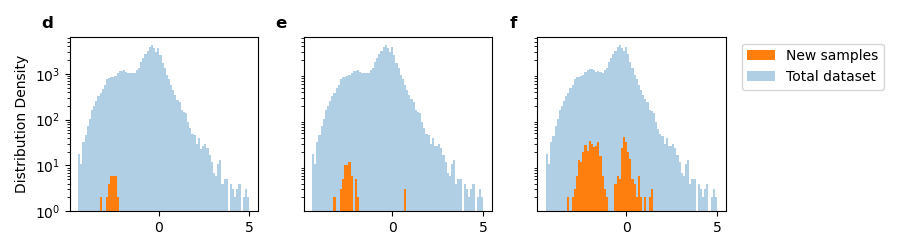}
    \end{subfigure}
    \begin{subfigure}[b]{\textwidth}
        \includegraphics[width=\textwidth]{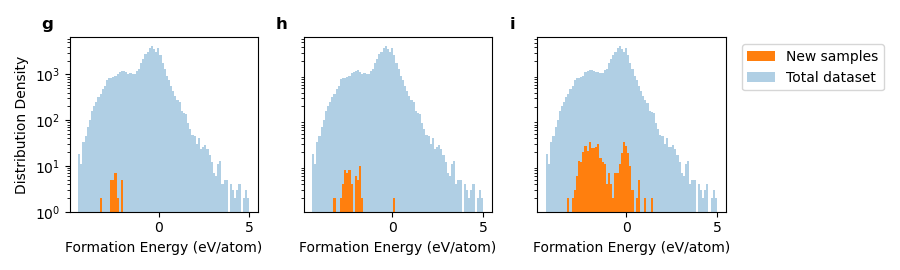}
    \end{subfigure}
    \caption{Using loss landscape uncertainties for active learning the formation energy omit Fe task. (a, b, c) Random sampling from uncalibrated loss landscapes. (d, e, f) Uncertainty calibrated using neural network. Column 1: 5~\% of the held out training data. Column 2: 10~\% of the held out training data. Column 3: 70~\% of the held out training data.}
    \label{fig:active_learning_dist_eform_Fe_ll}
\end{figure}

\begin{figure}
    \centering
    \includegraphics[width=\linewidth]{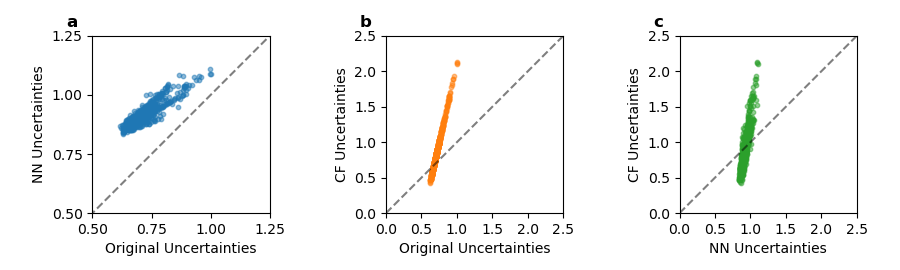}
    \caption{Comparison of how calibration methods transform uncertainties from the loss landscape ensemble trained for formation energy omit Fe task. (a) The values calibrated with the neural network, (b) values calibrated with calibration factors, and (c) a comparison of the calibration factors to the neural network calibration. The dashed line indicates where the two transforms are equivalent.}
    \label{fig:sample_ll_eform_Fe_comp_uncert}
\end{figure}

\begin{figure}[!h]
    \centering
    \begin{subfigure}[b]{\textwidth}
        \includegraphics[width=\textwidth]{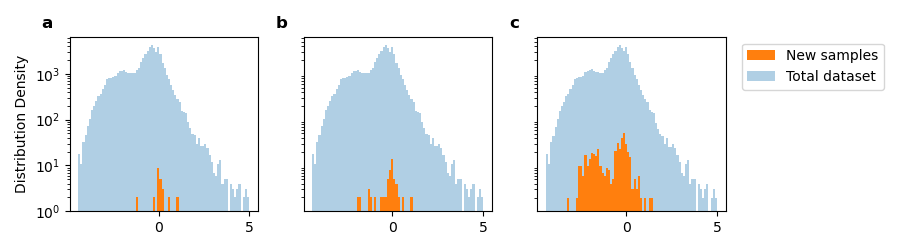}
    \end{subfigure}
    \begin{subfigure}[b]{\textwidth}
        \includegraphics[width=\textwidth]{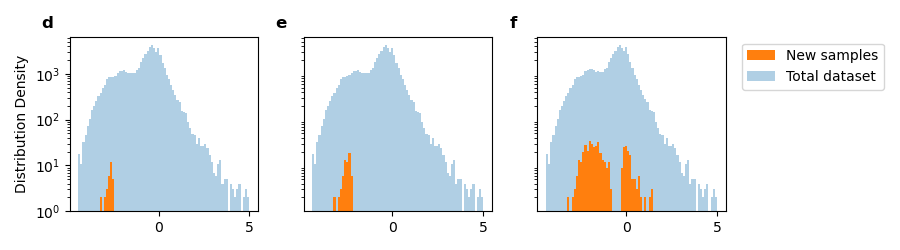}
    \end{subfigure}
    \begin{subfigure}[b]{\textwidth}
        \includegraphics[width=\textwidth]{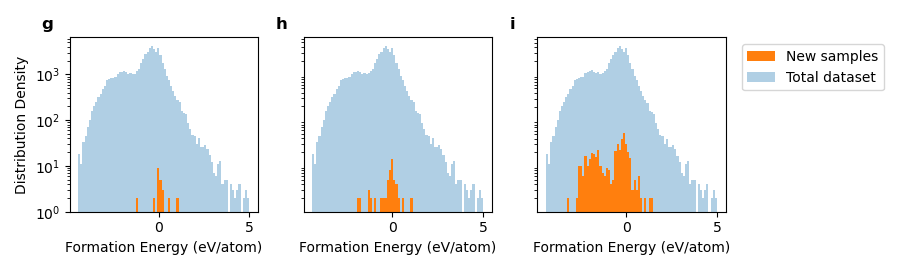}
    \end{subfigure}
    \caption{Using random ensemble uncertainties for active learning the formation energy omit Fe task. (a, b, c) Random sampling from uncalibrated loss landscapes. (d, e, f) Uncertainty calibrated using neural network. Column 1: 5~\% of the held out training data. Column 2: 10~\% of the held out training data. Column 3: 70~\% of the held out training data.}
    \label{fig:active_learning_dist_eform_Fe_ensem}
\end{figure}

\begin{figure}
    \centering
    \includegraphics[width=\linewidth]{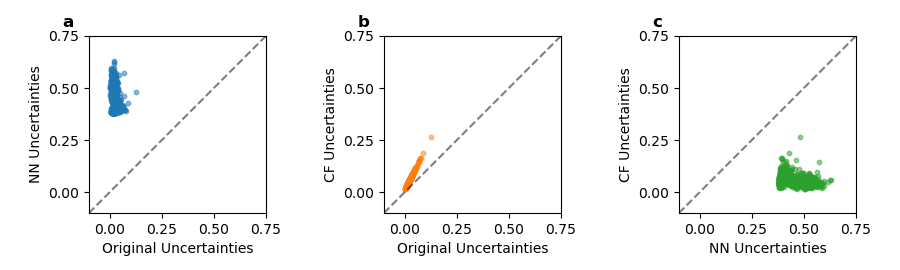}
    \caption{Comparison of how calibration methods transform uncertainties from the random ensemble trained for formation energy omit Fe task. (a) The values calibrated with the neural network, (b) values calibrated with calibration factors, and (c) a comparison of the calibration factors to the neural network calibration. The dashed line indicates where the two transforms are equivalent.}
    \label{fig:sample_ensem_eform_Fe_comp_uncert}
\end{figure}

\begin{figure}[!h]
    \centering
    \begin{subfigure}[b]{\textwidth}
        \includegraphics[width=\textwidth]{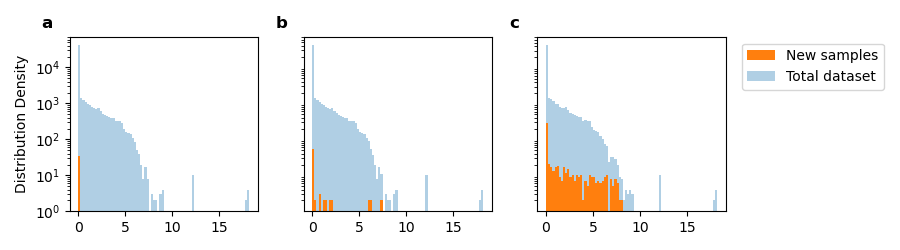}
    \end{subfigure}
    \begin{subfigure}[b]{\textwidth}
        \includegraphics[width=\textwidth]{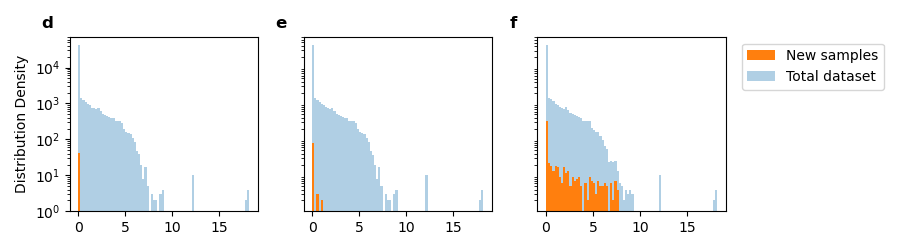}
    \end{subfigure}
    \begin{subfigure}[b]{\textwidth}
        \includegraphics[width=\textwidth]{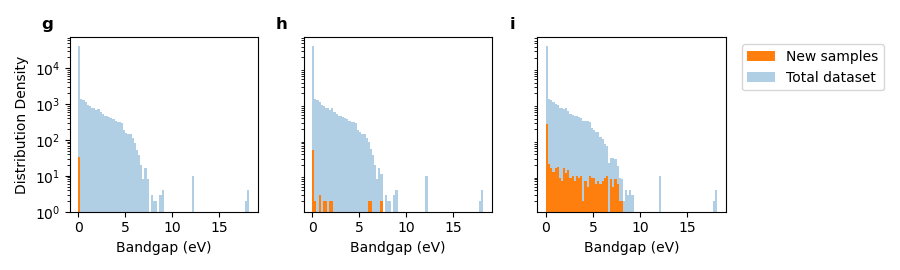}
    \end{subfigure}
    \caption{Using loss landscape uncertainties for active learning the bandgap omit F task. (a, b, c) Random sampling from uncalibrated loss landscapes. (d, e, f) Uncertainty calibrated using neural network. Column 1: 5~\% of the held out training data. Column 2: 10~\% of the held out training data. Column 3: 70~\% of the held out training data.}
    \label{fig:active_learning_dist_bandgap_F_ll}
\end{figure}

\begin{figure}
    \centering
    \includegraphics[width=\linewidth]{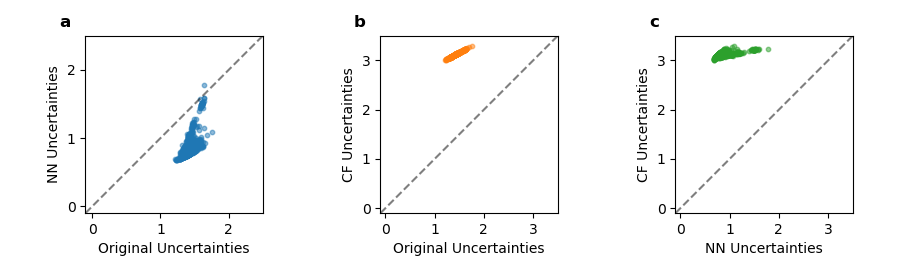}
    \caption{Comparison of how calibration methods transform uncertainties from the loss landscape ensemble trained for bandgap omit F task. (a) The values calibrated with the neural network, (b) values calibrated with calibration factors, and (c) a comparison of the calibration factors to the neural network calibration. The dashed line indicates where the two transforms are equivalent.}
    \label{fig:sample_ll_bgap_F_comp_uncert}
\end{figure}

\begin{figure}[!h]
    \centering
    \begin{subfigure}[b]{\textwidth}
        \includegraphics[width=\textwidth]{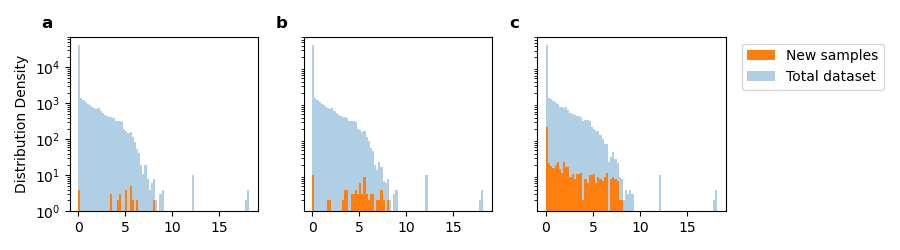}
    \end{subfigure}
    \begin{subfigure}[b]{\textwidth}
        \includegraphics[width=\textwidth]{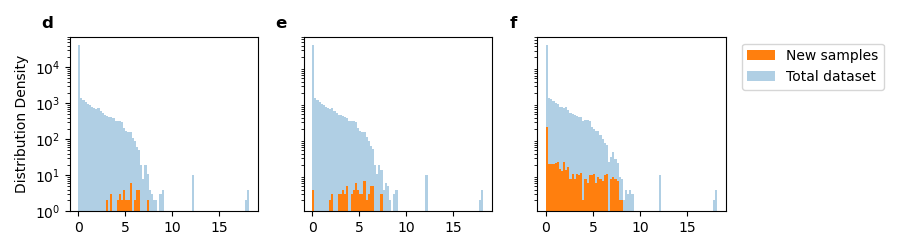}
    \end{subfigure}
    \begin{subfigure}[b]{\textwidth}
        \includegraphics[width=\textwidth]{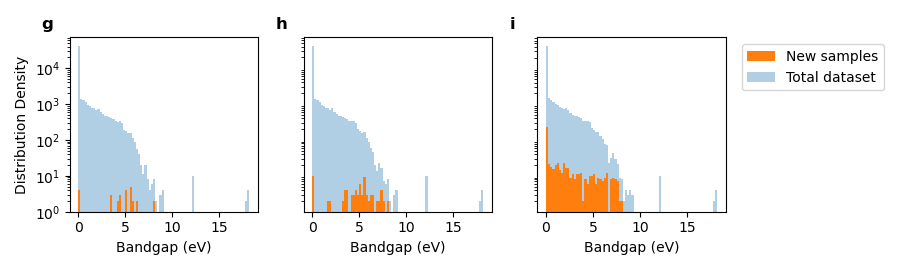}
    \end{subfigure}
    \caption{Using random ensemble uncertainties for active learning the bandgap omit F task. (a, b, c) Random sampling from uncalibrated loss landscapes. (d, e, f) Uncertainty calibrated using neural network. Column 1: 5~\% of the held out training data. Column 2: 10~\% of the held out training data. Column 3: 70~\% of the held out training data.}
    \label{fig:active_learning_dist_bandgap_F_ensem}
\end{figure}

\begin{figure}
    \centering
    \includegraphics[width=\linewidth]{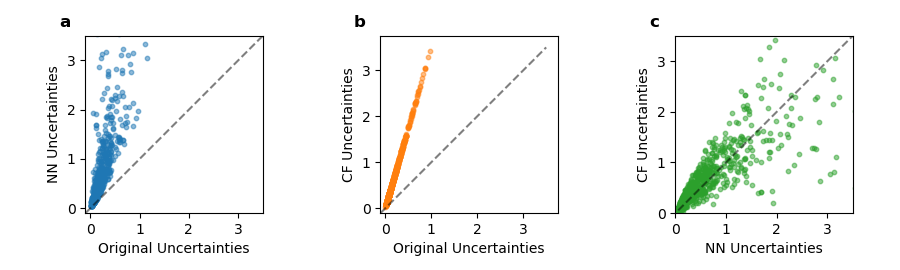}
    \caption{Comparison of how calibration methods transform uncertainties from the random ensemble trained for bandgap omit F task. (a) The values calibrated with the neural network, (b) values calibrated with calibration factors, and (c) a comparison of the calibration factors to the neural network calibration. The dashed line indicates where the two transforms are equivalent.}
    \label{fig:sample_ensem_bgap_F_comp_uncert}
\end{figure}

\begin{figure}[!h]
    \centering
    \begin{subfigure}[b]{\textwidth}
        \includegraphics[width=\textwidth]{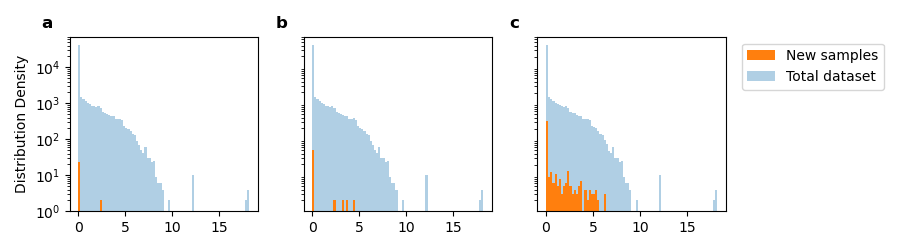}
    \end{subfigure}
    \begin{subfigure}[b]{\textwidth}
        \includegraphics[width=\textwidth]{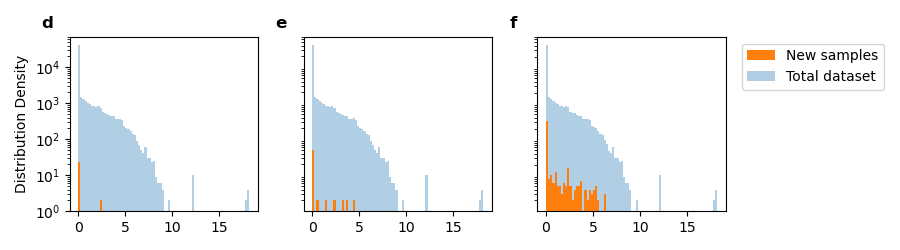}
    \end{subfigure}
    \begin{subfigure}[b]{\textwidth}
        \includegraphics[width=\textwidth]{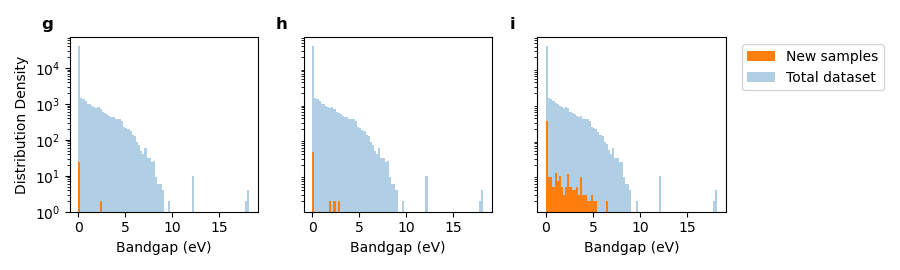}
    \end{subfigure}
    \caption{Using loss landscape uncertainties for active learning the bandgap omit Fe task. (a, b, c) Random sampling from uncalibrated loss landscapes. (d, e, f) Uncertainty calibrated using neural network. Column 1: 5~\% of the held out training data. Column 2: 10~\% of the held out training data. Column 3: 70~\% of the held out training data.}
    \label{fig:active_learning_dist_bandgap_Fe_ll}
\end{figure}

\begin{figure}
    \centering
    \includegraphics[width=\linewidth]{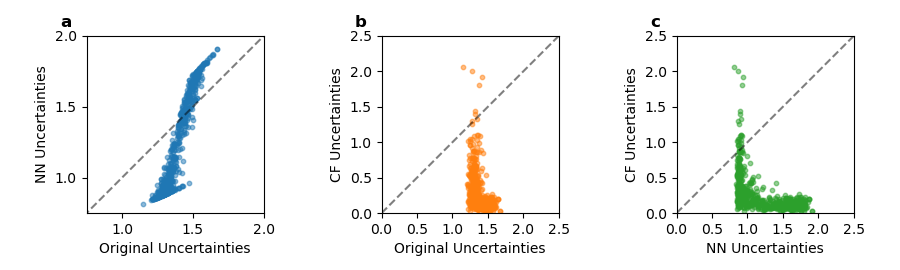}
    \caption{Comparison of how calibration methods transform uncertainties from the loss landscape ensemble trained for bandgap omit Fe task. (a) The values calibrated with the neural network, (b) values calibrated with calibration factors, and (c) a comparison of the calibration factors to the neural network calibration. The dashed line indicates where the two transforms are equivalent.}
    \label{fig:sample_ll_bgap_Fe_comp_uncert}
\end{figure}

\begin{figure}[!h]
    \centering
    \begin{subfigure}[b]{\textwidth}
        \includegraphics[width=\textwidth]{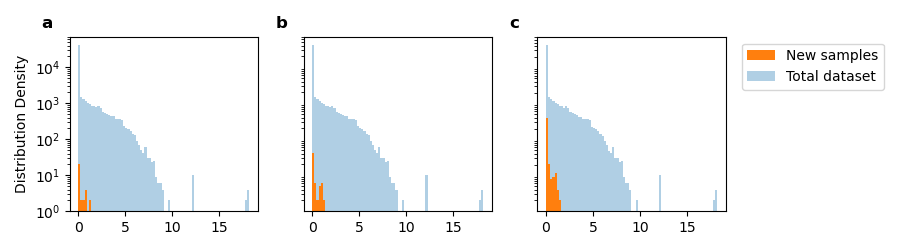}
    \end{subfigure}
    \begin{subfigure}[b]{\textwidth}
        \includegraphics[width=\textwidth]{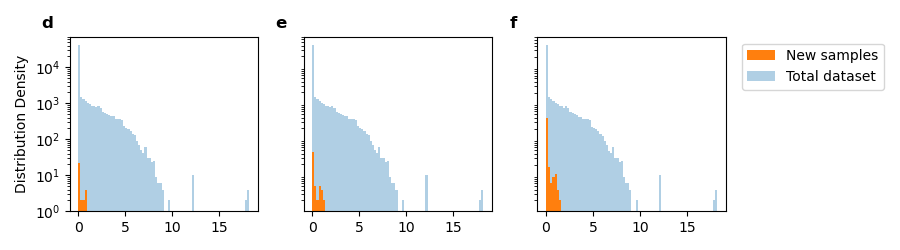}
    \end{subfigure}
    \begin{subfigure}[b]{\textwidth}
        \includegraphics[width=\textwidth]{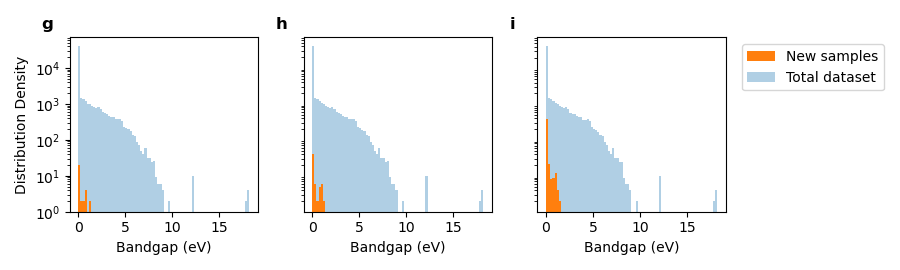}
    \end{subfigure}
    \caption{Using random ensemble uncertainties for active learning the bandgap omit Fe task. (a, b, c) Random sampling from uncalibrated loss landscapes. (d, e, f) Uncertainty calibrated using neural network. Column 1: 5~\% of the held out training data. Column 2: 10~\% of the held out training data. Column 3: 70~\% of the held out training data.}
    \label{fig:active_learning_dist_bandgap_Fe_ensem}
\end{figure}

\begin{figure}
    \centering
    \includegraphics[width=\linewidth]{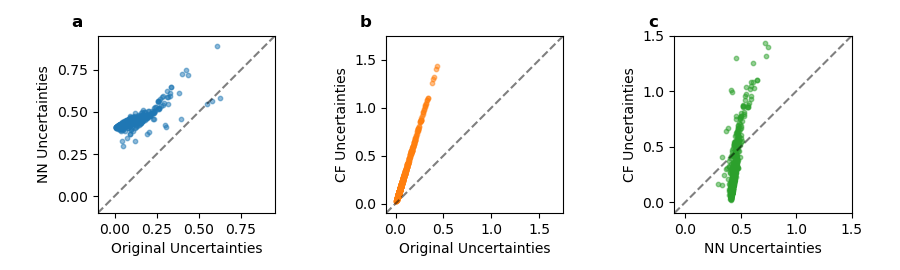}
    \caption{Comparison of how calibration methods transform uncertainties from the random ensemble trained for bandgap omit Fe task. (a) The values calibrated with the neural network, (b) values calibrated with calibration factors, and (c) a comparison of the calibration factors to the neural network calibration. The dashed line indicates where the two transforms are equivalent.}
    \label{fig:sample_ensem_bgap_Fe_comp_uncert}
\end{figure}
\end{appendices}
\end{document}